\documentclass[]{aastex631}
\usepackage{CJK}
\usepackage{multirow}

\received{April 21, 2023}
\revised{June 20, 2023}
\accepted{June 20, 2023}

\submitjournal{AJ}

\shorttitle{Gap Stars Activities}
\shortauthors{Jao et al.}
\graphicspath{{./}{figures/}}

\begin{document}

\title{Mind the Gap I: H$\alpha$ Activity of M Dwarfs Near the Partially/Fully Convective Boundary and a New H$\alpha$ Emission Deficiency Zone on the Main Sequence}

\correspondingauthor{Wei-Chun Jao}
\email{wjao@gsu.edu}

\author[0000-0003-0193-2187]{Wei-Chun Jao}
\affil{Department of Physics and Astronomy \\
Georgia State University \\
Atlanta, GA 30303, USA}

\author[0000-0002-9061-2865]{Todd J. Henry}
\affil{RECONS Institute, Chambersburg, PA 17201, USA}

\author[0000-0001-5313-7498]{Russel J. White}
\affil{Department of Physics and Astronomy \\
Georgia State University \\
Atlanta, GA 30303, USA}

\author[0000-0002-1457-1467]{Azmain H. Nisak}
\affil{Astronomy Department \\
Wesleyan University \\
Middletown, CT 06459, USA}

\author[0000-0003-4568-2079]{Hodari-Sadiki Hubbard-James}
\affil{Department of Physics and Astronomy \\
Georgia State University \\
Atlanta, GA 30303, USA}

\author[0000-0003-1324-0495]{Leonardo A. Paredes}
\affil{Department of Physics and Astronomy \\
Georgia State University \\
Atlanta, GA 30303, USA} 
\affil{Department of Astronomy/Steward Observatory \\
University of Arizona \\
Tucson, AZ 85721, USA
}

\author{Vanders B. Lewis, Jr.}
\affil{Department of Physics and Astronomy \\
Georgia State University \\
Atlanta, GA 30303, USA}

\begin{abstract}
 Since identifying the gap in the H-R Diagram (HRD) marking the transition between partially and fully convective interiors, a unique type of slowly pulsating M dwarf has been proposed. These unstable M dwarfs provide new laboratories in which to understand how changing interior structures can produce potentially observable activity at the surface. In this work, we report the results of the largest high-resolution spectroscopic H$\alpha$ emission survey to date spanning this transition region, including 480 M dwarfs observed using the CHIRON spectrograph at CTIO/SMARTS 1.5-m. We find that M dwarfs with H$\alpha$ in emission are almost entirely found 0 to 0.5 magnitude above the top edge of the gap in the HRD, whereas effectively no stars in and below the gap show emission.  Thus, the top edge of the gap marks a relatively sharp activity transition, and there is no anomalous H$\alpha$ activity for stars in the gap.  We also identify a new region at 10.3 $<M_{G}<$ 10.8 on the main sequence where fewer M dwarfs exhibit H$\alpha$ emission compared to M dwarfs above and below this magnitude range. Careful evaluation of literature results indicates that 1) rotation and H$\alpha$ activity distributions on the main sequence are closely related, and 2) fewer stars in this absolute magnitude range rotate in less than $\sim$13 days than populations surrounding this region.  This result suggests that the most massive fully convective stars lose their angular momentum faster than both partially convective stars and less massive fully convective stars.

\end{abstract}


\section{Introduction} 
\label{sec:intro}

The main sequence gap on the Hertzsprung--Russell diagram (HRD) reported by \cite{Jao2018} marks the interior structure transition between partially convective and fully convective interiors in M dwarfs. Stars above the gap are more massive and have an interior like the Sun, with a radiative zone and a convection layer, resulting in an $\alpha\Omega$ magnetic dynamo \citep{Charbonneau2014}. Stars below the gap are less massive, have fully convective interiors, and a different $\alpha^2$ dynamo \citep{Chabrier2006}. After the announcement of the discovery of the gap, \cite{MacDonald2018}, \cite{Baraffe2018}, and \cite{Feiden2021} found that stars in the gap with masses between 0.34 and 0.36 M$_{\odot}$ experience a period of structural instability and radial pulsations caused by non-equilibrium $^3$He fusion in the stellar core; this results in intermittent mixing between the core and envelope convection zones. Those efforts match earlier theoretical work by \cite{Saders2012}, who dubbed the resulting process the ``convective kissing instability’’. This suggests that the interiors of stars in the gap are more complex than stars above and below the gap, and may have up to three layers of energy transport: a convective core with an overlying radiative zone, which is in turn surrounded by a convective envelope. The underlying physics of this $^3$He instability drives slow pulsations in stellar radii and luminosities that are thought to produce the gap in the HRD. \cite{Mansfield2021} showed the effects of these oscillations on the evolutionary tracks of these gap stars on the HRD. The overall result of these modeling efforts indicates that the low mass stars in the gap are likely unique on the main sequence.

Stellar dynamos are driven by differential rotation and convection, but several questions arise when considering unstable stars in the gap: What stellar dynamos occur for stars having not one, but two layers of convective energy transport that shift over time? Do they have ``switching’’ magnetic dynamos, changing between $\alpha\Omega$ and $\alpha^2$, depending on the stage along their stellar evolutionary paths?  Do they exhibit enhanced activity at the surface beyond typical rotation-induced activity? Although detailed theoretical work has not yet been done to understand the complete ramifications of these unstable interiors, and whether or not they trigger enhanced surface activity, some circumstantial evidence has been presented.

The tachocline is the boundary between the radiative interior and the outer convective zone. \cite{Gilman2005} argued that the tachocline plays a significant role in creating magnetic patterns observed in photospheres, which consequently results in spot activity. \cite{Donati2008} and \cite{Reiners2009} showed that stars with masses close to 0.34--0.36 $M_{\odot}$ are likely to experience abrupt changes in their large-scale magnetic topologies, which can also trigger activity signatures at the surface. Hence, it is possible, perhaps even likely, that stars in the gap could have different photometric and spectroscopic characteristics from stars on either side of the gap because of their unique and changing interior structures. 

In this first of a series of papers entitled ``Mind the Gap'', we target stars close to the narrow transition boundary between partially and fully convective M dwarfs to evaluate their spectroscopic and photometric properties. Here we present new results of high-resolution spectroscopic observations for 480 stars close to the gap to understand their H$\alpha$ emission/absorption behavior.  We also evaluate complementary data available in the literature for a broader context. For our spectroscopic survey, we outline target selection in $\S$\ref{sec:targets}, followed by details of data acquisition and reduction in $\S$\ref{sec:observation}. Our assessment of these results is presented in $\S$\ref{sec:results}. We analyze results from four previous studies that include large numbers of M dwarf H$\alpha$ and magnetic measurements in $\S$\ref{sec:others}. Finally, in $\S$\ref{sec:newfeature}, we report on a new feature where we find a dearth of active M dwarfs on the main sequence and make a first attempt to understand the cause of this H$\alpha$ emission deficiency.  We present our conclusions in $\S$\ref{sec:conclusion}.

\section{Target Selection}
\label{sec:targets}

\begin{figure}
    \centering
    \plotone{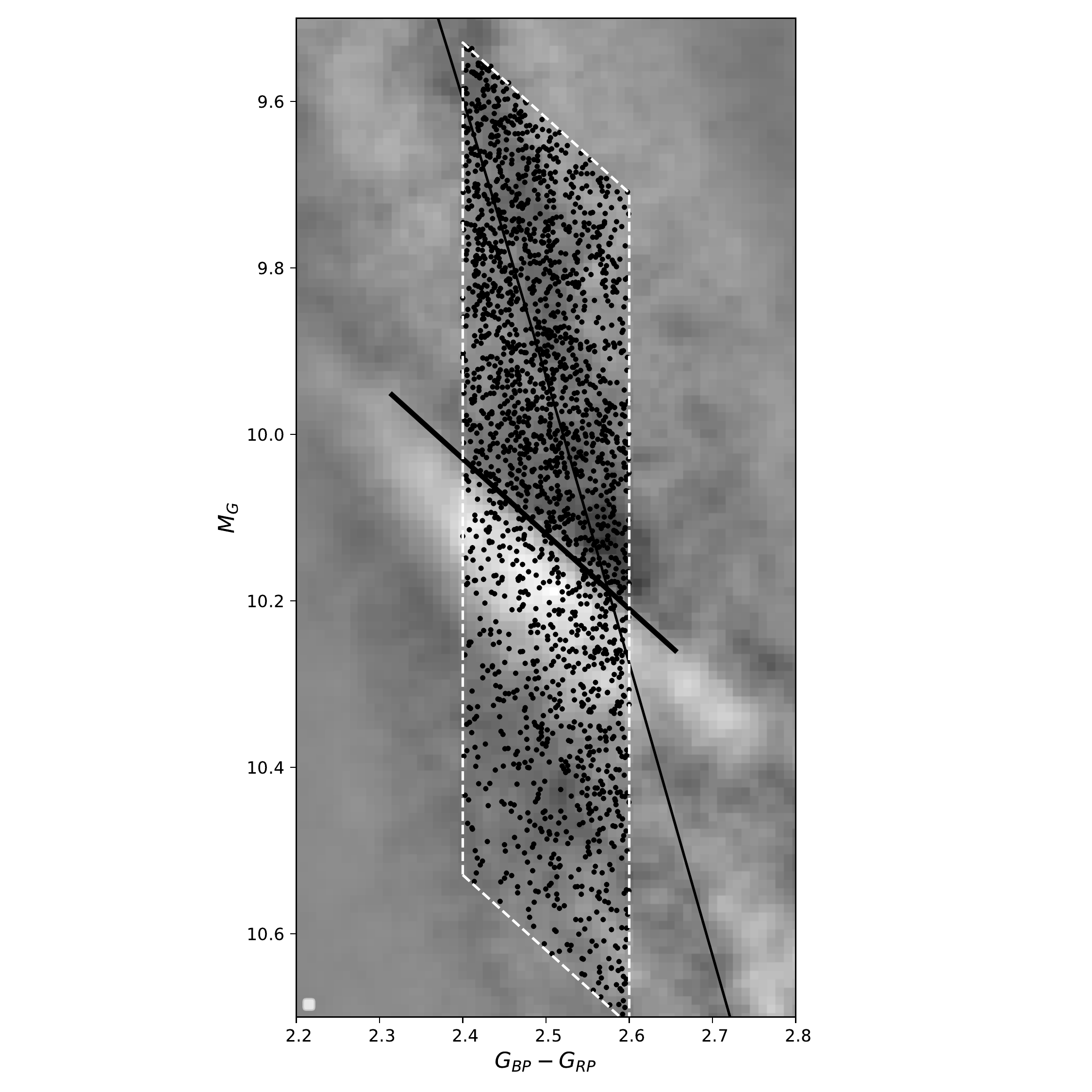}
    \caption{The sample of 2,249 M dwarf targets (black dots) meeting our selection criteria are shown in a small region of the HRD \citep{Jao2020}, where the thin black line traces the best-fit main sequence. The white shading is the gap first reported in \cite{Jao2018}, enhanced here to show where the stellar density is lower than expected. The thick black line marks the high-luminosity gap edge, hereafter referred to as ``GE''.  Details of defining the main sequence and GE lines are given in Appendix~\ref{sec:appendex1}. The tilted white region of the gap across the main sequence drives our selection of the white-dashed parallelogram that is the region interest (ROI); this permits evaluation of subsamples in incremental offsets above and below the GE. The same parallelogram ROI is used throughout this work.}
    \label{fig:ROI}
\end{figure}

The main sequence gap in the HRD is narrow, with a width of at most $\sim$0.2 in $M_G$ and is tilted relative to the main sequence \citep{Jao2020}. The gap is not evident across the entire width of the main sequence, fading at the blue and red ends where fewer stars are available for plotting. To select targets close to this fine feature for this work, we focus on stars (1) with $G_{BP}-G_{RP}$ color between 2.4 and 2.6 (roughly spectral types M3.0V to M4.0V) and (2) within $\pm$0.5 mag of the gap, the upper edge of which is defined by a fitted thick black line shown in Figure~\ref{fig:ROI} and given in Table~\ref{tbl:coeffs}. This fitted line is known as the gap edge, hereafter called "GE". Our region of interest (ROI) is a parallelogram extending above and below the GE centered at the widest part of the gap at $G_{BP}-G_{RP}$ = 2.5. The narrow color range for our sample ensures that the selected stars have similar spectral energy distributions and effective temperatures, thereby minimizing effects that would come into play with a less homogeneous sample. The same parallelogram ROI is used throughout this paper and outlined in various Figures. 

For our spectroscopic survey, we selected presumably single stars in the ROI that have high-quality astrometric data in the Gaia Early Data Release 3 catalog \citep[EDR3,][]{EDR3}. We applied four criteria to select stars in fields that provide uncontaminated spectra so that any detected activity can be considered coming from the targets. In total, 2,249 stars met our criteria; the entire target list is available electronically in Table~\ref{tbl:2249} and shown in Figure~\ref{fig:ROI}. The selection criteria are:

\begin{enumerate}
    \item distances within 100 pc: nearby stars
    \item RUWE $<$ 1.4: likely single stars
    \item $|b| >$ 10: away from the Galactic plane
    \item no source in EDR3 with $\Delta G_{RP}<$ 4.0 within 63\arcsec~of the target, thereby providing minimal contamination within three pixels of the star in the TESS (Transiting Exoplanet Survey Satellite) camera

\end{enumerate}

\begin{table}[h]
   \caption{Selected M Dwarf Targets Close to the Gap}
    \centering
    \begin{tabular}{ccl}
    \hline
    \hline
    name & units & description \\
    \hline
    source\_id &  & source ID in Gaia EDR3\\
    R.A. & deg & Right ascension in Gaia EDR3 epoch\\
    Decl. & deg & Declination in Gaia EDR3 epoch \\
    parallax & mas & parallax in Gaia EDR3\\
    pmra & mas & proper motion in R.A. in Gaia EDR3\\
    pmdec & mas & proper motion in Decl. in Gaia EDR3\\
    RUWE & & Renormalised Unit Weight Error\\
    G & mag & magnitude in Gaia EDR3 G band\\
    BP & mag & magnitude in Gaia EDR3 BP band \\
    RP & mag & magnitude in Gaia EDR3 RP band \\
    \hline
    \end{tabular}
    \tablecomments{This table is available online electronically.}
    \label{tbl:2249}
\end{table}

\section{Spectroscopic Observations and Data Reduction}
\label{sec:observation}

\subsection{CHIRON Spectroscopic Observations}

We observed 443 of the 2,249 stars using the CHIRON high-resolution spectrograph \citep{Tokovinin2013, Paredes2021} on the SMARTS 1.5-m telescope at Cerro Tololo Inter-American Observatory (CTIO). This is the largest high-resolution spectroscopic survey to date that focuses on the narrow transition zone between partially and fully convective low mass stars. Because some targets were observed before we completed the target selection, an additional 37 targets that don’t meet the 63\arcsec~contamination criterion are also included. None of these have a contaminating source within CHIRON’s fiber diameter of 2\farcs7, so in total, we observed and presented results for 480 stars. These stars have $G$ magnitudes of 8.2--13.9 and a mean parallax of 28 mas.

Observations were made at the 1.5-m between November 2018 and August 2022. Of the four different resolutions available with CHIRON, 96\% of the targets are observed in the fiber mode with R = 28,000 that covers 4500-8300\AA~over 62 spectral orders. The remaining targets were observed using slicer mode with R = 79,000 covering the same wavelengths over 59 orders. All targets have estimated $V$ magnitudes brighter than 15, so exposure times of 900--1800 seconds were used, depending on brightness, to reach a mean S/N of 23. A ThAr lamp image was taken immediately after each target for an exposure of 0.25 or 4 seconds to calibrate the wavelength for fiber and slicer modes, respectively. We carried out standard data processing procedures that included bias and flat field corrections, cosmic-ray removal, echelle order extractions, wavelength calibrations using ThAr comparison lamps, and removal of the blaze function. The details of data processing are fully discussed in \citet{Paredes2021}.

For this work, the key feature is the H$\alpha$ line found at 6562.8\AA~in air in the rest frame.  To pinpoint the feature, we do a barycentric correction using the algorithm discussed in \citet{Wright2014} and provided via their online tool\footnote{\url{https://astroutils.astronomy.osu.edu/exofast/barycorr.html}}. The radial velocity (RV) of the star is then calculated based on a spectral cross-correlation against Barnard’s star (M4.0V, \citealt{Henry1994}) and GJ 273 (M3.5V, \citealt{Henry1994}), for which we adopt radial velocities of $-$110.11 km s$^{-1}$$\pm$0.2 m s$^{-1}$ and 18.36 km s$^{-1}$$\pm$0.3 m s$^{-1}$, respectively from \cite{Fouque2018}. To obtain RVs, a cross-correlation analysis is performed as described in \citet{Irwin2018} and \cite{Nisak2022}, using six different orders in fiber mode: order 38 (6397.94---6471.62\AA), 39 (6471.47---6546.00\AA), 41 (6623.73---6699.99\AA), 42 (6702.58---6779.74\AA), 46 (7037.69---7118.67\AA), and 53 (7712.50---7801.18\AA). Similar wavelength coverage is used for data obtained in the slicer mode, but the echelle orders are shifted two orders higher. The RV of a target in each of the six orders is calculated relative to both of the RV reference stars.  After offsetting for the standard stars' velocities, the two sets of 6 values are then used to determine the final weighted mean RV and error for each target. In total, we acquired 503 spectra of 480 targets, where twenty-one targets have more than one epoch, and found a mean RV error of 1.8 km s$^{-1}$.  The RV measurements, as well as results for H$\alpha$ equivalent width and $v\sin i$ discussed next, are given for our targets in Table~\ref{tbl:results}.

\subsection{Determining H$\alpha$ Equivalent Widths}
\label{sec:defineEW}

After making the barycentric correction and consequent shift of the spectrum to the rest frame for a given star, we measure the equivalent width (EW) of the H$\alpha$ line to measure the star's activity level. Measuring an EW value is accomplished via the equation

\begin{equation}
    EW=\sum \left ( 1-\frac{F_{mean}(\lambda)}{F_{C}} \right ) \Delta\lambda
\end{equation}

\noindent where $F_{mean}(\lambda)$ is the flux between F1 and F2 for the feature, 
$F_{C}$ is the mean pseudo-continuum flux, and $\Delta\lambda$ is 0.023\AA~for slicer mode data or for 0.094\AA~fiber mode data.  Here we adopt the usual convention with negative EW values for stars with H$\alpha$ in emission and positive EW values for stars with H$\alpha$ in absorption. We calculate EW errors using the relation given in \cite{Cayrel1988}. 

\begin{figure}
    \centering
    \includegraphics[scale=0.7]{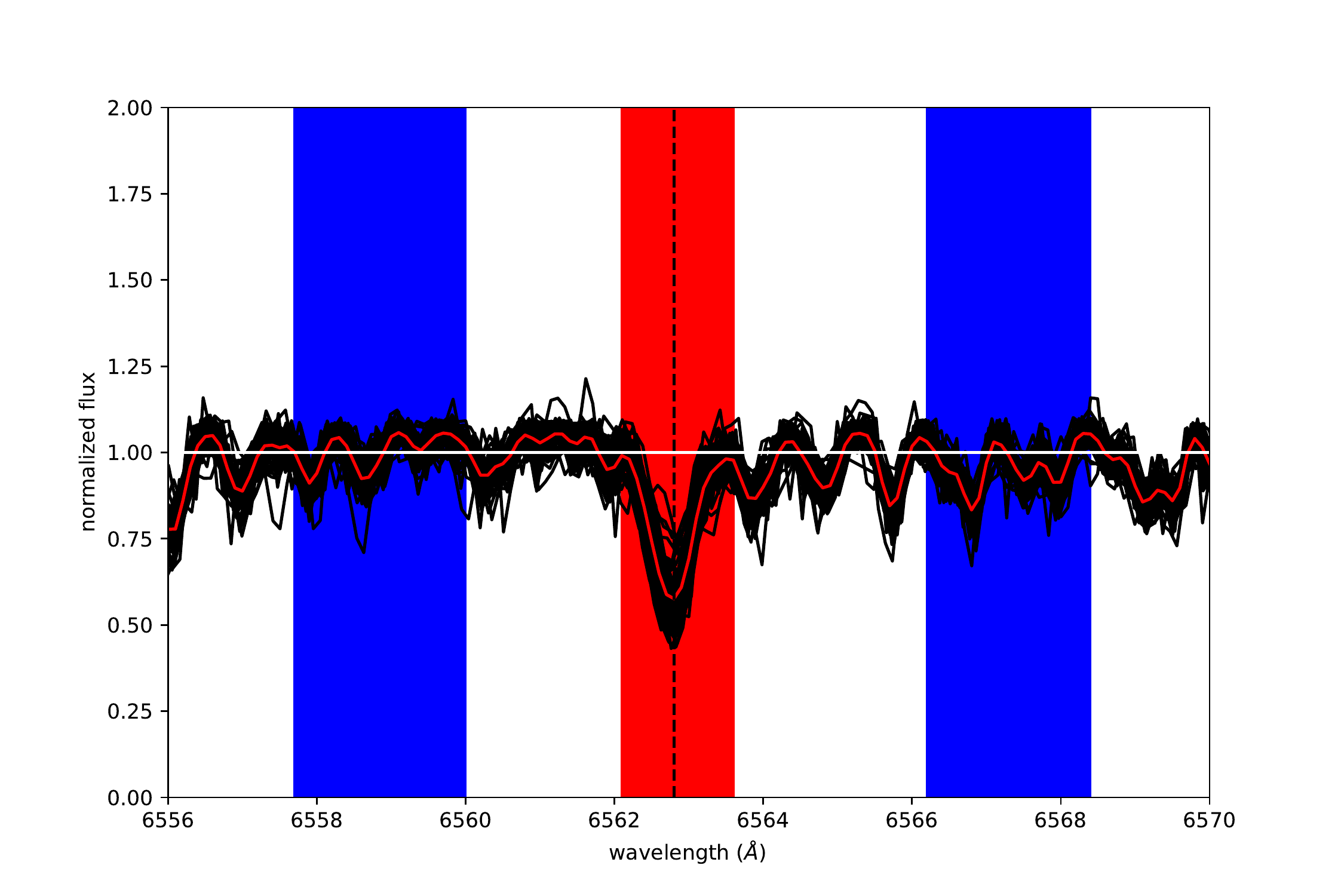}
    \caption{Windows used to determine H$\alpha$ EW values. The black lines trace spectra for 79 stars with H$\alpha$ absorption features and the red line represents the mean spectrum for these stars. The red box is the nominal spectral window used to measure the H$\alpha$ line, with the dashed line marking the feature at 6562.8\AA~in air.  The two blue boxes mark the windows for the pseudo-continuum on either side.  Wavelength values for all three window edges are given in Table~\ref{tbl:EWs}.}
    \label{fig:EWtemplate}
\end{figure}

To measure the H$\alpha$ EW values, a window centered on the feature and two windows on either side representing the pseudo-continuum were defined.  We randomly selected 79 stars observed in fiber mode with well-defined H$\alpha$ absorption features as templates to define the windows, illustrated with overlapping black lines in Figure~\ref{fig:EWtemplate}.  The same windows are used for stars observed in the slicer mode. The wavelengths of the window edges are given in Table~\ref{tbl:EWs}; these windows are similar to those used in \citet{Gizis2002}. For all stars with emission except for GAI1439$+$1023, which has an emission fitted within the default template, the central H$\alpha$ line window was extended to include flux from the wings by visual comparison to the template. In these cases, the windows of the pseudo-continuum on either side were shifted outward to exclude flux from the strong emission, while maintaining the window widths.

\begin{deluxetable}{ccccccc}
\tablecaption{H$\alpha$ line\label{tbl:EWs}}
\tablehead{
\colhead{} &
\multicolumn{2}{c}{left continuum} &
\multicolumn{2}{c}{adopted line} &
\multicolumn{2}{c}{right continuum}\\
\hline
\colhead{} &
\colhead{C1} &
\colhead{C2} &
\colhead{F1} &
\colhead{F2} &
\colhead{C3} &
\colhead{C4}
}
\startdata
H$\alpha$ & 6557.7 & 6560.0 & 6562.1 & 6563.6 & 6566.2 & 6568.4 \\
\enddata
\tablecomments{Units are Angstroms.}
\end{deluxetable}

We note here that the H$\alpha$ equivalent width is often converted to the ratio of H$\alpha$ luminosity and bolometric luminosity to account for the changing continuum across the wide range of temperatures exhibited by M dwarfs \citep{West2008, Newton2017, Kiman2021}.  Even without obtaining flux calibrated spectra, calculations can be done using the so-called $\chi$ factor \citep{Walkowicz2004, Douglas2014} to obtain $L_{\alpha}/L_{bol}$ values, where $L_{\alpha}/L_{bol}= -EW \times \chi$. This conversion is needed for the analysis of M dwarf samples spanning temperatures corresponding to a color range from early to late M dwarfs of $\sim$4 magnitudes in $G_{BP}-G_{RP}$.  However, the color range of our sample is only a very narrow 0.2 magnitude in $G_{BP}-G_{RP}$, so conversions are not necessary.  In fact, converting EWs to $L_{\alpha}/L_{bol}$ would have errors from this relation itself and introduce additional errors because the commonly used relations to obtain $\chi$ use $V-I_c$ or $i-J$ colors rather than the {\it Gaia} photometry used here.  Consequently, the equivalent width presented in this work can be interpreted as a direct tracer of H$\alpha$ luminosity without any conversions.

\subsection{Projected Rotational Velocities}

We also measured the projected rotational velocities ($v\sin i$) for the survey stars using the prescription described in \cite{Nisak2022}. Before we measured the $v\sin i$, we first established an empirical relation between the width of the cross-correlation function (CCF) and $v\sin i$ by cross-correlating each standard star spectrum against rotationally broadened synthetic versions of itself using the rotational broadening routine in PyAstronomy\footnote{\url{https://github.com/sczesla/PyAstronomy}}.  The CCF width of a survey star is calculated from all six orders of the spectrum used to determine its RV. Then, using the mean CCF width from the six orders and the empirical relation, a mean $v\sin i$ is derived. As a check, three previously known fast rotating stars were observed with CHIRON and measured using this method, with our results and those from earlier work given in Table~\ref{tbl:fast}; in eight cases out of nine, our values are consistent with previously published values within the errors, with the single outlier that of GJ871.1A, for which we suspect the value of \cite{Fouque2018} is offset from the true value.  Given the spectral resolution of our observations, S/N of the spectra, and the standard stars we use, we conservatively only report stars with $v\sin i$ faster than 10 km/s in fiber mode and 4 km/s in slicer mode. In total, only seven stars in fiber mode have measurable rotation signatures, and no star in slicer mode has $v\sin i$ faster than the threshold. Results are given in Table~\ref{tbl:results}.

\begin{deluxetable}{lrrcrcrc}
\tablecaption{Known fast rotation stars\label{tbl:fast}}
\tablehead{
\colhead{Target} &
\colhead{This work} &
\colhead{\#1} &
\colhead{Ref1}&
\colhead{\#2} &
\colhead{Ref2}&
\colhead{\#3} &
\colhead{Ref3}\\
\hline
\colhead{} &
\multicolumn{7}{c}{$v\sin i$ (km/s)}
}
\startdata
TWA9B         &  7.08(1.79) &   9.0(1.0) & 1 &   8.39(0.61) & 2  & 10.9(1.7) & 6 \\
2MA0200-0840  & 15.13(1.75) &  12.2(2.1) & 5 &  16.1(2.1)   & 4  & 15.2(1.4) & 4 \\
GJ871.1A      & 12.34(1.36) &  20.5(2.4) & 5 &  14.5(1.9)   & 4  & 13.9(0.5) & 3 \\
\enddata
\tablecomments{All results are in fiber modes. The $v \sin i$ errors are within parentheses.}
\tablerefs{1. \cite{White2004}, 2. \cite{Scholz2007}, 3. \cite{Bailey2012}, 4. \cite{Malo2014}, 5. \cite{Fouque2018}, 6. \cite{Lopez2021}}
\end{deluxetable}

\begin{deluxetable}{ccccccccccccccccccccc}
\tablecaption{Results\label{tbl:results}}
\tabletypesize{\tiny}
\rotate
\tablehead{
\colhead{source ID} &
\colhead{Name} & 
\colhead{R.A.} & 
\colhead{Decl.} & 
\colhead{$\pi$} & 
\colhead{RUWE} & 
\colhead{$G$} & 
\colhead{$G_{BP}-G_{RP}$} & 
\colhead{neighbor} &
\colhead{\# visit} &
\colhead{decker} & 
\colhead{epoch} & 
\colhead{activity} & 
\colhead{RV} & 
\colhead{RV$_{err}$} & 
\colhead{$v\sin i$} & 
\colhead{$v\sin i_{err}$} & 
\colhead{EW} & 
\colhead{EW$_{err}$} &
\colhead{CaH1} &
\colhead{binary}\\
\colhead{} &
\colhead{} & 
\colhead{deg} & 
\colhead{deg} & 
\colhead{mas} & 
\colhead{} & 
\colhead{mag} & 
\colhead{mag} & 
\colhead{} &
\colhead{} & 
\colhead{} & 
\colhead{years} & 
\colhead{} & 
\colhead{km s$^{-1}$} & 
\colhead{km s$^{-1}$} & 
\colhead{km s$^{-1}$} & 
\colhead{km s$^{-1}$} & 
\colhead{\AA} & 
\colhead{\AA} &
\colhead{} &
\colhead{}
\\
\colhead{(1)} & 
\colhead{(2)} & 
\colhead{3)} & 
\colhead{(4)} & 
\colhead{(5)} & 
\colhead{(6)} & 
\colhead{(7)} & 
\colhead{(8)} & 
\colhead{(9)} & 
\colhead{(10)} & 
\colhead{(11)} & 
\colhead{(12)} & 
\colhead{(13)} & 
\colhead{(14)} & 
\colhead{(15)} & 
\colhead{(16)} &
\colhead{(17)} &
\colhead{(18)} &
\colhead{(19)} &
\colhead{(20)} &
\colhead{(21)}
}
\startdata
  2336284229875185024 & GAI0007-2524 & 00:07:06.39 & -25:24:51.65 & 23.9587 & 1.28 & 13.718 & 2.595 & 0 & 1 & fiber & 2021.6584 & A & 12.98 & 0.07 &  &  & 0.31 & 0.02 & 0.458 & \\
  4901109094112344192 & GAI0008-6401 & 00:08:26.69 & -64:01:09.09 & 30.4698 & 1.23 & 12.244 & 2.433 & 0 & 1 & fiber & 2021.5929 & A & -9.83 & 0.18 &  &  & 0.37 & 0.02 & 0.628 & \\
  4991695142706265344 & GAI0012-4445 & 00:12:03.04 & -44:45:06.80 & 13.9347 & 1.15 & 13.938 & 2.444 & 0 & 1 & fiber & 2021.855 & A & 16.12 & 0.22 &  &  & 0.36 & 0.02 & 0.483 & \\
  4901473650937640576 & GAI0012-6312 & 00:12:26.65 & -63:12:43.62 & 19.6865 & 1.16 & 13.928 & 2.449 & 0 & 1 & fiber & 2021.8551 & A & 46.85 & 0.16 &  &  & 0.3 & 0.02 & 0.729 & \\
  4706483132131647232 & GAI0015-6759 & 00:15:45.06 & -67:59:38.22 & 53.2872 & 1.36 & 11.447 & 2.447 & 1 & 1 & slicer & 2018.92 & A & -30.79 & 0.02 &  &  & 0.24 & 0.01 & 0.573 & \\
  2309132133824849152 & GAI0018-3450 & 00:18:44.75 & -34:50:19.47 & 42.033 & 1.38 & 11.791 & 2.577 & 0 & 1 & fiber & 2021.5956 & A & 24.89 & 0.06 &  &  & 0.34 & 0.02 & 0.559 & \\
  2360419059860637184 & GAI0022-2230 & 00:22:24.06 & -22:30:42.37 & 17.3802 & 1.35 & 13.557 & 2.453 & 0 & 1 & fiber & 2021.6368 & A & 25.27 & 0.07 &  &  & 0.29 & 0.02 & 0.647 & \\
  2423826899002038912 & GAI0027-1203 & 00:27:28.44 & -12:03:10.32 & 19.5732 & 1.19 & 13.763 & 2.421 & 0 & 1 & fiber & 2021.6885 & A & -30.56 & 0.26 &  &  & 0.19 & 0.02 & 0.444 & \\
  2528082144871347584 & GAI0028-0429 & 00:28:10.12 & -04:29:12.07 & 18.6548 & 1.16 & 13.716 & 2.556 & 0 & 1 & fiber & 2021.6175 & A & -10.93 & 0.3 &  &  & 0.2 & 0.02 & 0.566 & \\
  2367344127690031744 & GAI0028-1700 & 00:28:17.67 & -17:00:59.72 & 18.6458 & 1.27 & 13.345 & 2.494 & 0 & 1 & fiber & 2021.8634 & A & -18.12 & 0.66 &  &  & 0.25 & 0.01 & 0.687 & 1\\
\enddata
\tablecomments{Column (9): Stars meeting our selection criteria discussed in Table~\ref{tbl:2249} have a code of 0; 37 additional stars have a code of 1. Column (10): number of observations. Column (11): spectra were taken with CHIRON in fiber or slicer mode. Column (12): epoch of RV measurement. Column (13): ``A'' indicates H$\alpha$ in absorption and ``E'' indicates in emission. Column (16): projected rotational velocities. Column (21): Possible binaries. The full table is available electronically online.}
\end{deluxetable}

\section{Results}
\label{sec:results}

\begin{figure}
    \centering
    \plotone{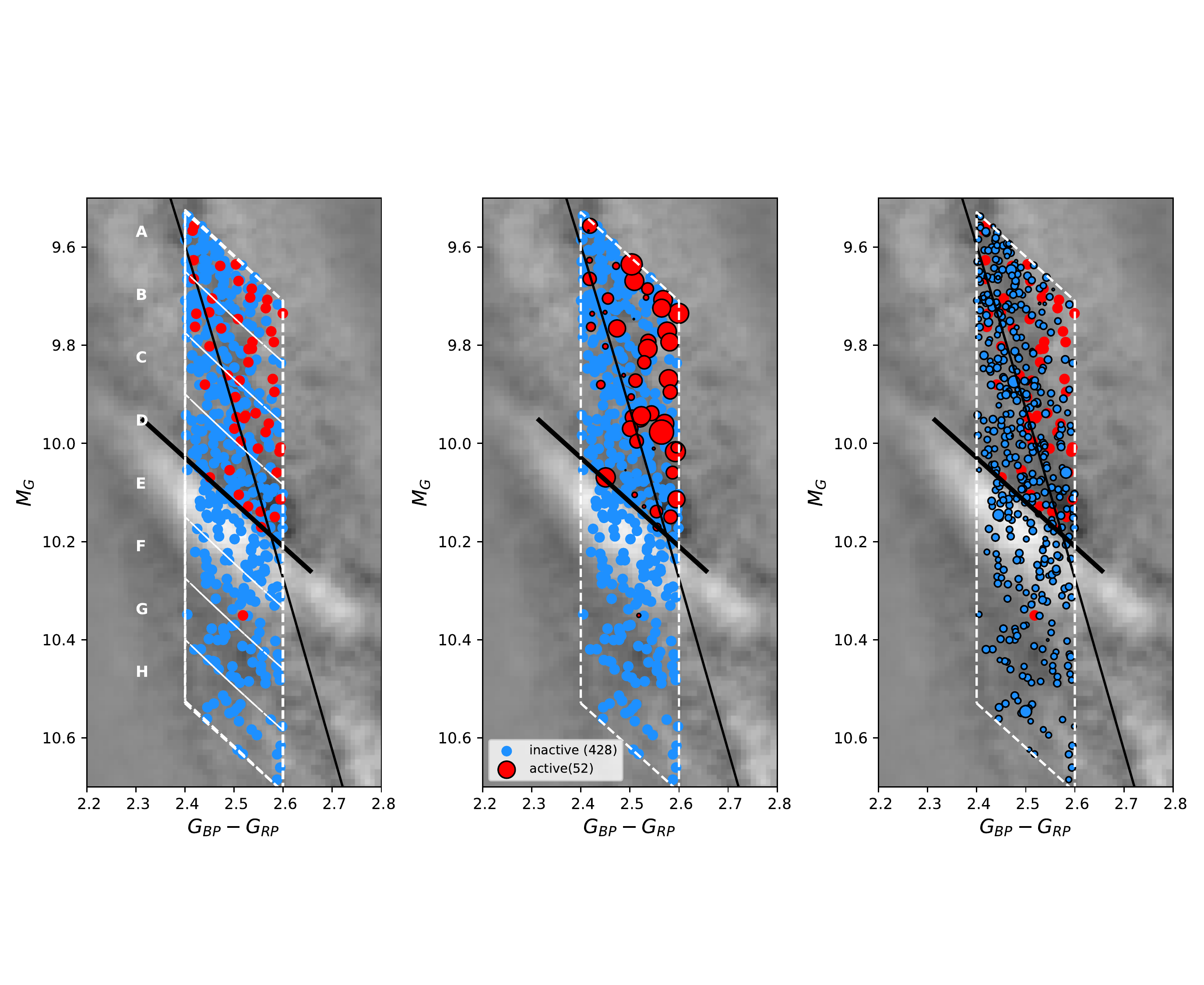}
    \caption{(Left): Active stars (red dots) and inactive stars (blue dots) on the enhanced HRD, identified using H$\alpha$ equivalent widths (EW) in CHIRON spectra. Eight equal size zones are outlined inside the ROI. The gap edge (GE) is shown with a short heavy black line , and the best-fit main sequence line is shown with a long thin black line. (Center): The absolute values of EWs for active stars are plotted as a function of the size of red circles. The single active star represented by the very small red circle below the GE is GAI0406-5011 with a very low EW$=-0.18$\AA. (Right): The EWs of inactive stars are indicated by the size of blue circles; all have similar EWs.}
    \label{fig:results}
\end{figure}

\subsection{Activity Distribution Across the Gap}
\label{sec:acrossgap}

Previous studies of M dwarf H$\alpha$ activity have often sorted samples by spectral types, colors, or estimated masses \citep[e.g.,][]
{Gizis2002, West2008, Walkowicz2009}, and samples were analyzed using one-dimensional graphs. However, to study H$\alpha$ activity across the gap, a two-dimensional feature, we plot active and inactive targets on the two-dimensional HRD shown in Figure~\ref{fig:results} so that we can understand how both colors and luminosities relate to H$\alpha$ activity, and map activity relative to the gap and the across the main sequence. This figure shows that effectively all active stars identified by this work, shown with red points, are above the GE. The only active star (GAI0406-5011) clearly below the GE has weak H$\alpha$ emission, with an EW$=-0.18$\AA; this star appears to be single based on Gaia catalog and the CCF peak. Our result suggests that effectively all active stars in our survey are partially convective, whereas fully convective stars show no H$\alpha$ emission. As a result, instead of finding an H$\alpha$ activity anomaly {\it in the gap}, we find that the activity distribution is sharply divided at the GE.  In other words, even though stars in the gap have changing interior structures and slowly pulsate, their H$\alpha$ activity does not differ from other fully convective stars. In addition, most active stars are found above the best-fit main sequence line in Figure~\ref{fig:results}.

To further investigate the H$\alpha$ emission distribution for our targets, we divide the ROI into eight equal-sized smaller zones. As shown in Figure~\ref{fig:results}, we outline four zones (A--D) above and four below (E--H) the GE, with the width of each zone approximately the width of the gap, which corresponds to zone E. The percentages of active stars in each zone are given in Table~\ref{tbl:percentage}, which shows that the number of active stars clearly drops below the GE. In zones A--D, there are 50 active stars out of 334, indicating that 15\% exhibit H$\alpha$ emission.  In zones E--H, there are only 2 active stars out of 146, or 1\%. There is only one star (GAI0516-3846) in zone E that is active, but that star is barely in zone E, found only 0.002 mags below the GE. Given the magnitude error, it may be above the line. If we exclude this target, there is only one active star (GAI0406-5011) in the bottom half of the ROI, as shown in Figure~\ref{fig:results}. More discussions about this sharp transition are discussed in section~\ref{sec:conclusion}.

\begin{deluxetable}{cccccc|cc|cc}
\colnumbers
\tablecaption{Statistics for Samples of Stars in 8 Zones Near the Main Sequence Gap\label{tbl:percentage}}
\tablehead{
\multicolumn{1}{c}{} &
\multicolumn{5}{c|}{Sample Numbers} &
\multicolumn{2}{c|}{Mean CaH1 EWs} &
\multicolumn{2}{c}{Mean Galactic Velocities}\\
\hline
\colhead{Zone} &
\colhead{Total Stars} &
\colhead{Active Stars} &
\colhead{Inactive Stars} &
\colhead{\% Active} &
\colhead{\% Active} &
\colhead{$\langle$CaH1$\rangle $} &
\colhead{$\sigma_{CaH1}$} &
\colhead{$\langle$T$_{total}$$\rangle $} &
\colhead{$\sigma_{V}$}\\
\colhead{}&
\colhead{\#} &
\colhead{\#} &
\colhead{\#} &
\colhead{} &
\colhead{} &
\colhead{\AA} &
\colhead{\AA} &
\colhead{$km~sec^{-1}$} &
\colhead{$km~sec^{-1}$}
}
\startdata
A & 87 & 15 & 72 & 17.2$^{+4.8}_{-3.2}$ & \multirow{4}{*}{15.0$\pm$2.3} &  0.52 & 0.10 & 42.6 & 23.2\\
B & 80 & 13 & 67 & 16.2$^{+4.7}_{-3.2}$ &&  0.53 & 0.14 & 41.7 & 23.1\\
C & 85 & 15 & 70 & 17.6$^{+5.4}_{-3.6}$ &&  0.54 & 0.09 & 40.8 & 24.8\\
D & 82 &  7 & 75 &  8.5$^{+4.5}_{-2.5}$ &&  0.58 & 0.15 & 38.0 & 22.4\\
\hline
{\bf E (gap)} & {\bf 57} & {\bf(1)} & {\bf 56} & {\bf (1.8$^{+4.2}_{-0.8}$)} & \multirow{4}{*}{1.4$\pm$1.0} & {\bf 0.57} & {\bf 0.14} & {\bf 41.1} & {\bf 20.7}\\
F & 38 & 1 & 37 & 2.6$^{+5.4}_{-0.6}$ &&  0.58 & 0.18 & 41.4 & 20.6\\
G & 31 & 0 & 31 & 0                   &&  0.61 & 0.13 & 47.4 & 21.2\\
H & 20 & 0 & 20 & 0                   &&  0.58 & 0.09 & 42.5 & 24.1\\
\enddata
\tablecomments{The gap corresponds to zone E and is highlighted in bold. The parentheses indicate that this zone contains one star, only 0.002 mags below the line.  Quantities enclosed with $\langle \rangle$ are mean values in each zone. $T_{total}$ represents the a total Galactic velocity, $\sqrt{U^{2}+V^{2}+W^{2}}$.   $\sigma_{X}$ values indicate the standard deviations for CaH1 EW and $T_{total}$ in each zone. The percentage errors are calculated using the binomial distribution discussed in \cite{Burgasser2003} for a sample size of fewer than 100 stars in each bin.}

\end{deluxetable}

\subsection{Age Evaluations of the Observed Stars Via Metallicities and Kinematics}

Stellar activity in M dwarfs at a given color is often related to age because young stars are generally more active than old, low metallicity, inactive subdwarfs \citep{Jao2011, Clements2017}. Because all of our targets fall in a narrow color range, here we investigate whether or not the active stars in our sample are younger than the inactive stars by using stellar metallicity and kinematics as proxies for age.

The metallic molecular hydride bands, i.e., FeH, MgH, and CaH, are indicators of M dwarf metallicity \citep{Bessell1982, Gizis1997} and can be used to evaluate relative metallicities of stars rather than their absolute metallicities. For example, CaH band strength indices can be calculated to separate low metallicity subdwarfs from dwarfs, as discussed in \cite{Reid1995} and \cite{Gizis1997}. Here we use the CaH1 band at 6380--6390\AA~for the sample stars as a measurement of their relative metallicities.  Usually, the pseudo-continuum on either side of the molecular band is used to calculate the CaH1 index, and the lower the value determined for the index, the lower the metallicity. However, our spectra contain one side of the commonly used pseudo-continuum at 6345--6355\AA~defined in \cite{Gizis1997}, and the other side of the pseudo-continuum at 6410--6420\AA~is in a different echelle order. Our spectra are not flux calibrated, so stitching the orders is not straightforward. Instead of using continuum windows to calculate a CaH1 index, here we measure the EW of the CaH1 feature using the EW as defined in section~\ref{sec:defineEW}. A large EW means a deeper or stronger CaH1 band, which implies a larger ratio of line opacity to continuum opacity ($n(CaH1)/n(H^{-})$) or lower metallicity \citep{Bessell1982}.

\begin{figure}
    \centering
    \plotone{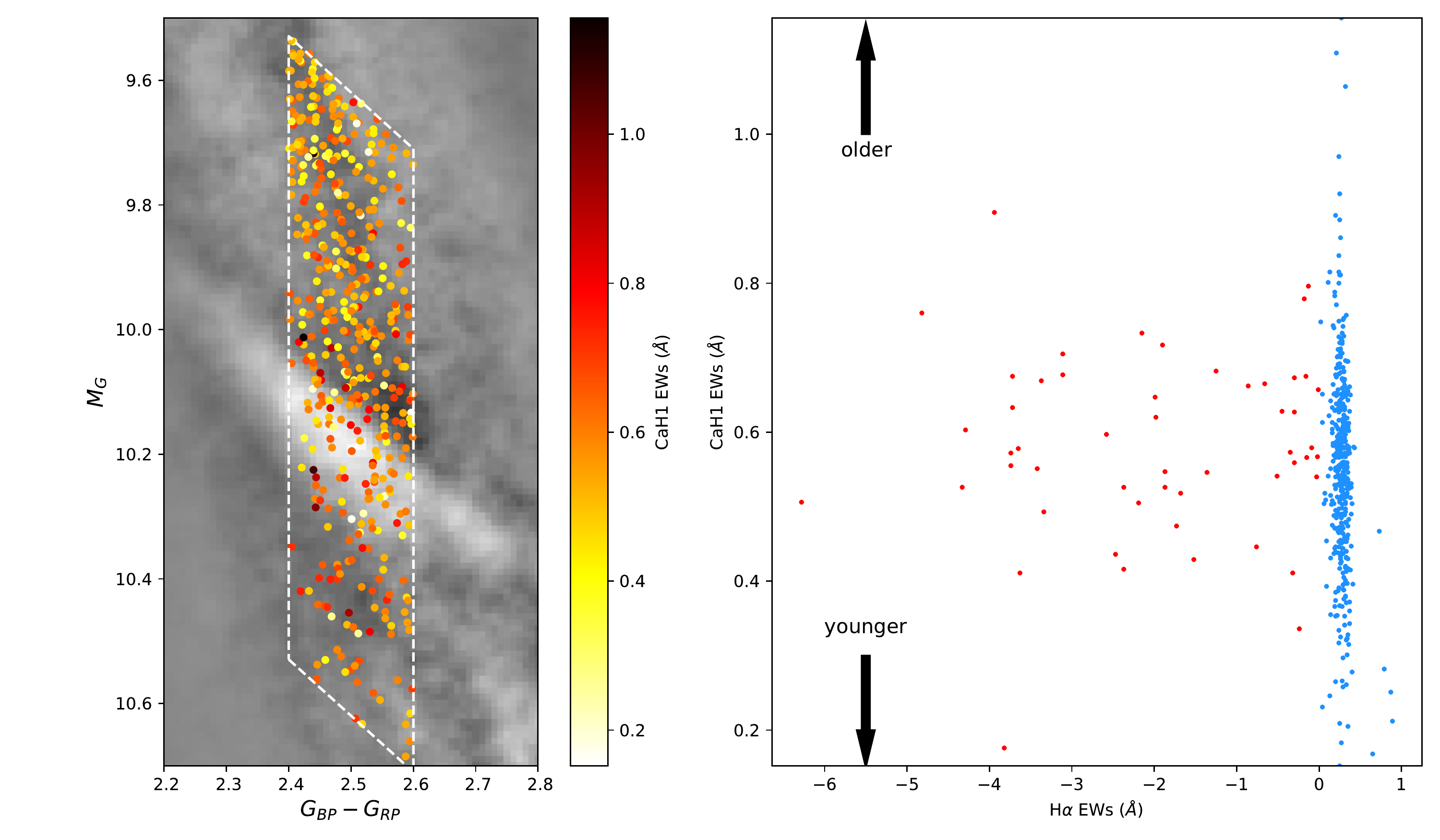}
    \caption{ (Left): The CaH band strength distribution for our targets in the enhanced HRD. This Figure is similar to Figure~\ref{fig:results}, but the two dark lines are omitted to show the contrast and distribution of the points. Smaller CaH1 EW values correspond to shallower CaH band strengths and higher metallicities. (Right) The EWs of H$\alpha$ and CaH1 features for active (red dots) and inactive stars (blue dots). The two arrows mark the directions of relatively older and younger stars.}
    \label{fig:CaH1}
\end{figure}

Figure~\ref{fig:CaH1} shows the CaH1 band strength distribution of our sample in the HRD, and Table~\ref{tbl:percentage} lists the mean CaH1 EWs in each region from A to H. There is a hint that there may be two metallicity groups --- one including the A to C regions with mean EWs of 0.52--0.54\AA~and the other the remaining regions from D to H with mean EWs of 0.57--0.61\AA. However, the standard deviations in the regions span 0.09--0.18, implying that the metallicity spread in the entire ROI is certainly not as distinct as the H$\alpha$ activity shown in Figure~\ref{fig:results}. In addition, the right panel of Figure~\ref{fig:results} shows that the active stars identified via H$\alpha$ emission have CaH1 EWs of $\sim$0.4--0.8\AA, whereas inactive stars span a wider range from $\sim$0.2--1.2\AA, with perhaps proportionally a few more stars at lower EW values. Overall, the CaH1 strengths for active and inactive stars do not differ significantly, which we interpret to mean that metallicity is not a significant factor affecting the activity distribution in Figure~\ref{fig:results} we see for H$\alpha$ in stars found in this narrow ROI.

\begin{figure}
    \centering
    \plotone{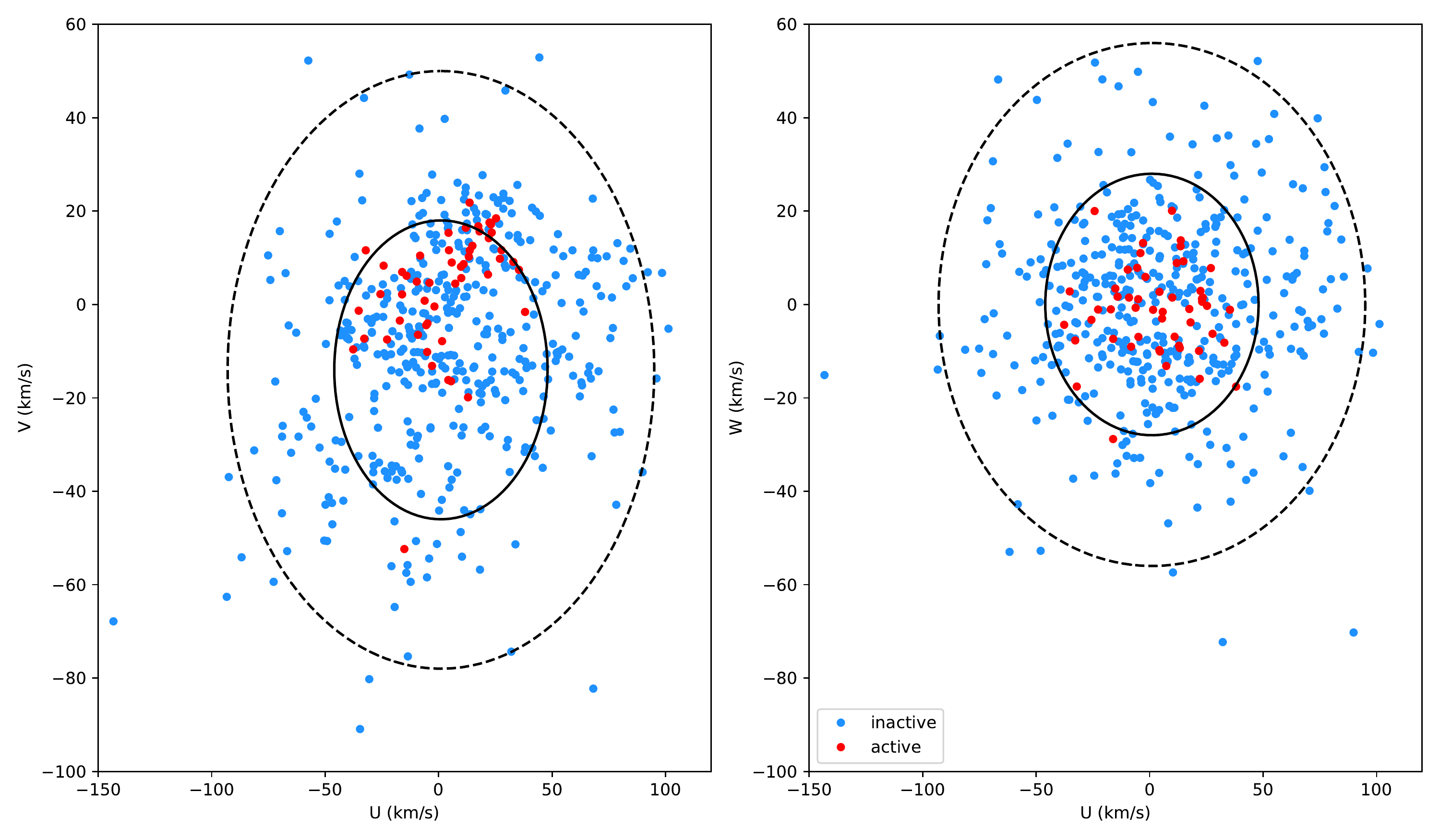}
    \caption{The Galactic UVW velocities of our targets are shown in two panels. The two black ellipses are centered on the UVW values of (1, 14, 0) km s$^{-1}$ for M dwarfs from \cite{Zhang2021b}. The ellipses trace 1$\sigma$ (solid line) and 2$\sigma$ (dotted line) regions of the dispersions with (47, 32, 28) km s$^{-1} $\citep{Zhang2021b}. Red dots are active stars, and blue dots are inactive stars. The location of mean UV values for cool subdwarfs (5, $-$248) in the left panel is off the left side of the plot. The velocities presented here are relative to the local standard of rest given in \cite{Schonrich2010}.}
    \label{fig:UVW}
\end{figure}

The other commonly used method to compare relative ages in a stellar sample is to evaluate their kinematics. The Galactic velocities of main sequence dwarfs and halo subdwarfs are different \citep{Hawley1996, Gizis1997, Zhang2021b} because, on average, low metallicity halo subdwarfs have faster Galactic velocities after billions of years of dynamic heating. In this work, we calculate the Galactic UVW velocities of our targets relative to the local standard of rest, for which the solar motion of (11.1, 12.24, 7.25) km s$^{-1}$ are adopted from \citep{Schonrich2010}. To calculate UVW values, coordinates, parallaxes, and proper motions are taken from {\it Gaia} DR3, and radial velocities are from this work. As shown in Figure~\ref{fig:UVW}, we find that the active stars are almost all within the solid line region marking 1$\sigma$ offsets from the mean velocity ellipsoid for M dwarfs, where we have adopted the mean UVW values of 
(1$\pm$47, 14$\pm$32, 0$\pm$28) km s$^{-1}$ from \cite{Zhang2021b}. Inactive stars, however, exhibit relatively wider velocity distributions, with nearly all falling inside the 2$\sigma$ offset region.  While the distribution is more dispersed for the inactive stars than for the active stars, the inactive stars are still much slower than the velocity ellipsoid of halo subdwarfs, which have mean UVW values of (5$\pm$173, -248$\pm$77, 6$\pm$89) km s$^{-1}$ \citep{Zhang2021b}. Kinematically, all of the stars in our sample appear to be thin or thick disk stars, and none appear to be halo subdwarfs. Table~\ref{tbl:percentage} presents the mean total Galactic velocities ($T_{total}=\sqrt{U^{2}+V^{2}+W^{2}}$) in each region, all of which fall in a narrow range of 38--48 km s$^{-1}$, with dispersions in each region of 20--25 km s$^{-1}$. As with the metallicity evaluation, there is no strong correlation between Galactic velocities and H$\alpha$ measurements, indicating that activity is not reflected in the kinematics across the various regions in the ROI.

\subsection{Spectroscopic Binaries}

Among the 480 stars in our sample, we identify 11 spectroscopic binary candidates in this work. Three stars show H$\alpha$ emission, and the remaining eight stars exhibit H$\alpha$ absorption features. All show more than one peak in the CCF when measuring radial and rotational velocities based on single spectroscopic observations, so they are presumably double-lined spectroscopic binaries with orbits that could be determined with additional observations. Both radial and rotational velocities for such systems are calculated based on the higher CCF peak in the reduction. Our targets were selected to have low RUWE ($<$1.4), and none of these stars are listed in the non-single stars list for {\it Gaia} DR3 \citep{nonsingle2022}. This implies they all have undetected astrometric perturbations through results derived for {\it Gaia} DR3, and that they are potentially nearly equal-mass binaries. These 11 systems are noted in Table~\ref{tbl:results}.

\section{Other Large Surveys Related to M Dwarf H$\alpha$ Activity}
\label{sec:others}

There are several large surveys of H$\alpha$ activity in M dwarfs in the literature, all of which span larger portions of the HRD compared to the tightly focused region explored here.  Here we consider, chronologically, four previous studies to understand H$\alpha$ activity for M dwarfs along the main sequence in order to place the H$\alpha$ activity for our sample in a broader context among larger populations of similar low mass stars.

\subsection{Newton 2017}

\citet{Newton2017}, hereafter N17, studied H$\alpha$ activity for 2,074 nearby M dwarfs with a mean parallax of 59.5 mas. After crossmatching their sample against {\it Gaia} EDR3 results to select stars with RUWE$<$1.4 and excluding targets noted as ``binary'' in their paper, there remain 1,249 M dwarfs that will be examined here, most of which are presumably single. Their reported H$\alpha$ EWs are from their own measurements and 27 earlier references; how they adopted EWs from various sources and calibrated the EWs are discussed in their work. They defined an H$\alpha$ EW threshold of $-$1\AA~as the dividing value between ``inactive'' and ``active'' stars, and an activity flag was assigned to each star in their Table 1. Only 12 stars are found in both N17 and our sample.  Of these, one star, GAI1019+1952, is flagged as active by both N17 and us, and the remaining 11 stars are identified as inactive.

\begin{figure}
\plotone{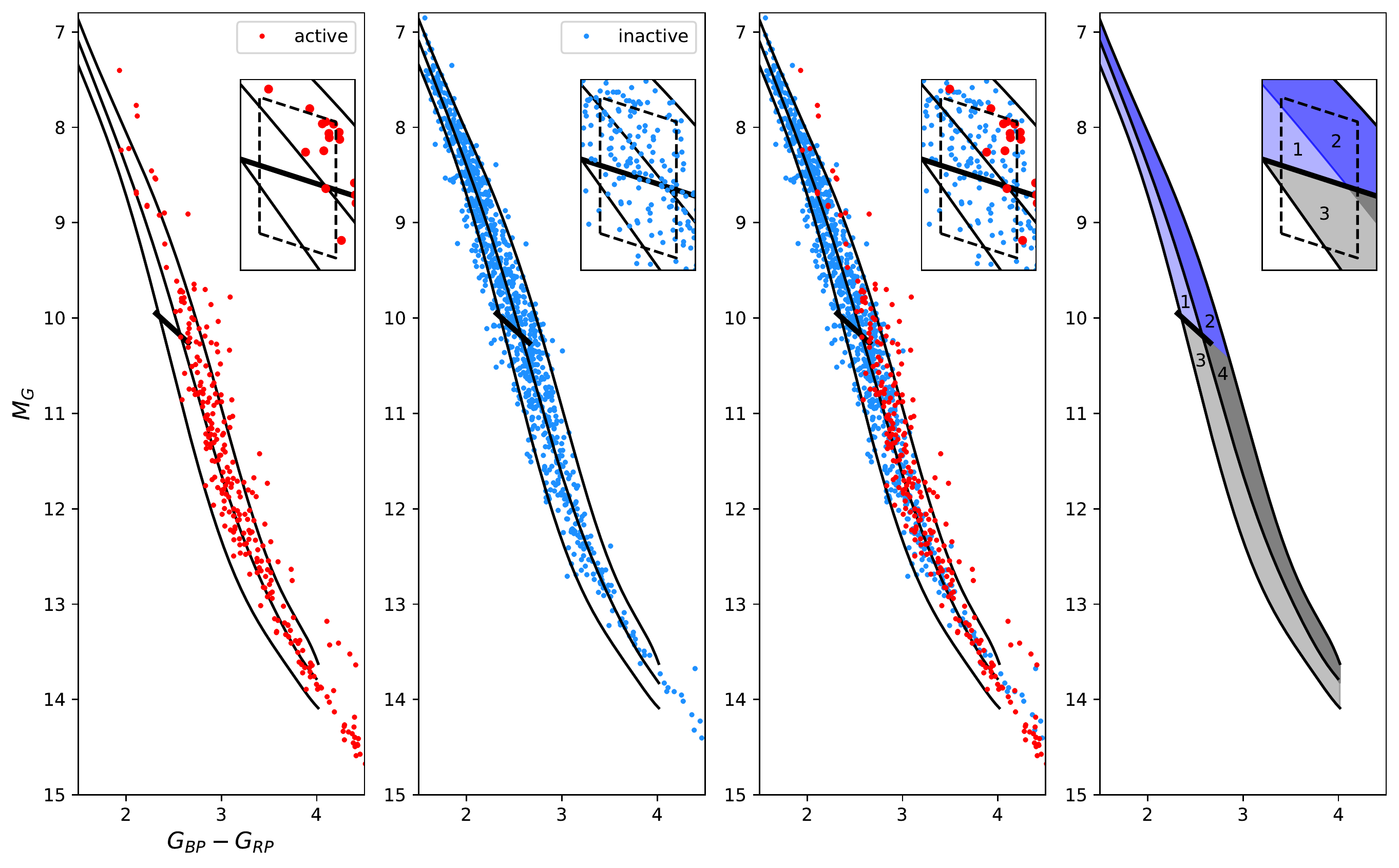}
\caption{The HRD of nearby M dwarfs from \citet{Newton2017}. The left plot is for M dwarfs flagged as ``active'' in \citet{Newton2017}, and the center-left plot is for ``inactive'' M dwarfs. The center-right plot merges the samples in the two left panels. The main sequence is divided into four regions, shown in the right plot. Black lines in these four plots represent the GE (short heavy line), the fitted main sequence, and the upper and lower envelopes of the main sequence. The polynomial coefficients of these fitted lines are given in Appendix~\ref{app:lines}. The insets highlight a region around our ROI, which is shown as a dashed parallelogram tracing the full ROI defined in Figure~\ref{fig:ROI}.}

\label{fig:Newton}
\end{figure}

The distributions of their active (red dots) and inactive (blue dots) stars are shown on the HRD in the first three panels of Figure~\ref{fig:Newton}. Because we study H$\alpha$ activity across this two-dimensional graph, we split the main sequence into four regions labeled 1 to 4 as shown in the full main sequence in the final, right panel of Figure~\ref{fig:Newton}. These regions are divided using polynomials so that there are two regions above (2 and 4) and two regions below (1 and 3) the best-fitted line of the main sequence. There are also two regions above (1 and 2) and two below (3 and 4) the GE. Details of generating and defining these lines are discussed in Appendix~\ref{app:lines}.

It is evident in the first panel of Figure~\ref{fig:Newton} that the majority of active M dwarfs in N17 are in regions 3 and 4, below the GE. Interestingly, most active stars lie above the median main sequence line, falling in regions 2 and 4. Some are even above the upper envelope of the main sequence, likely indicative of their youth because we have excluded known close binaries in this plot. The insets in Figure~\ref{fig:Newton} illustrate the region around the gap in the main sequence, with the heavy black line indicating the GE. Although their work contains over a thousand presumably single M dwarfs, there are only eight active stars falling inside our ROI. Seven of the eight are above the transition boundary, falling in the ROI's upper right corner, and only one active star is just below the transition line. The entire region below the GE is essentially void of active stars. The second panel of Figure~\ref{fig:Newton} shows that inactive stars are evenly distributed on either side of the fitted main sequence all along the main sequence, and the inset illustrates that some inactive stars are found above and below the GE.  Overall, the distributions of active and inactive stars within the ROI are similar in our work presented here and N17.

Next, we examine regions 3 and 4 below the GE. The ratio of inactive stars in regions 4 and 3 is 147:187 (0.79:1), and the ratio of active stars in regions 4 and 3 is 114:65 (1.75:1). Thus, there are proportionally many more active stars above the fitted main sequence than below. The broader perspective of M dwarf activity from N17 shown in Figure~\ref{fig:Newton} is that most active M dwarfs are fully convective, and in fact, there are no active stars found region 1, which is above the GE but below the median main sequence. This dearth of active stars in region 1 differs from our work and two other efforts discussed later in this section because all three efforts have identified active M dwarfs in region 1. Recall that the large sample reported in N17 results from an ensemble of 27 different efforts, so we suspect there may be biases due to sample selection or observations that have precluded the identification of active stars in region 1.

\subsection{Jeffers 2018}
\label{sec:CARMENES}

\begin{figure}
    \centering
    \plotone{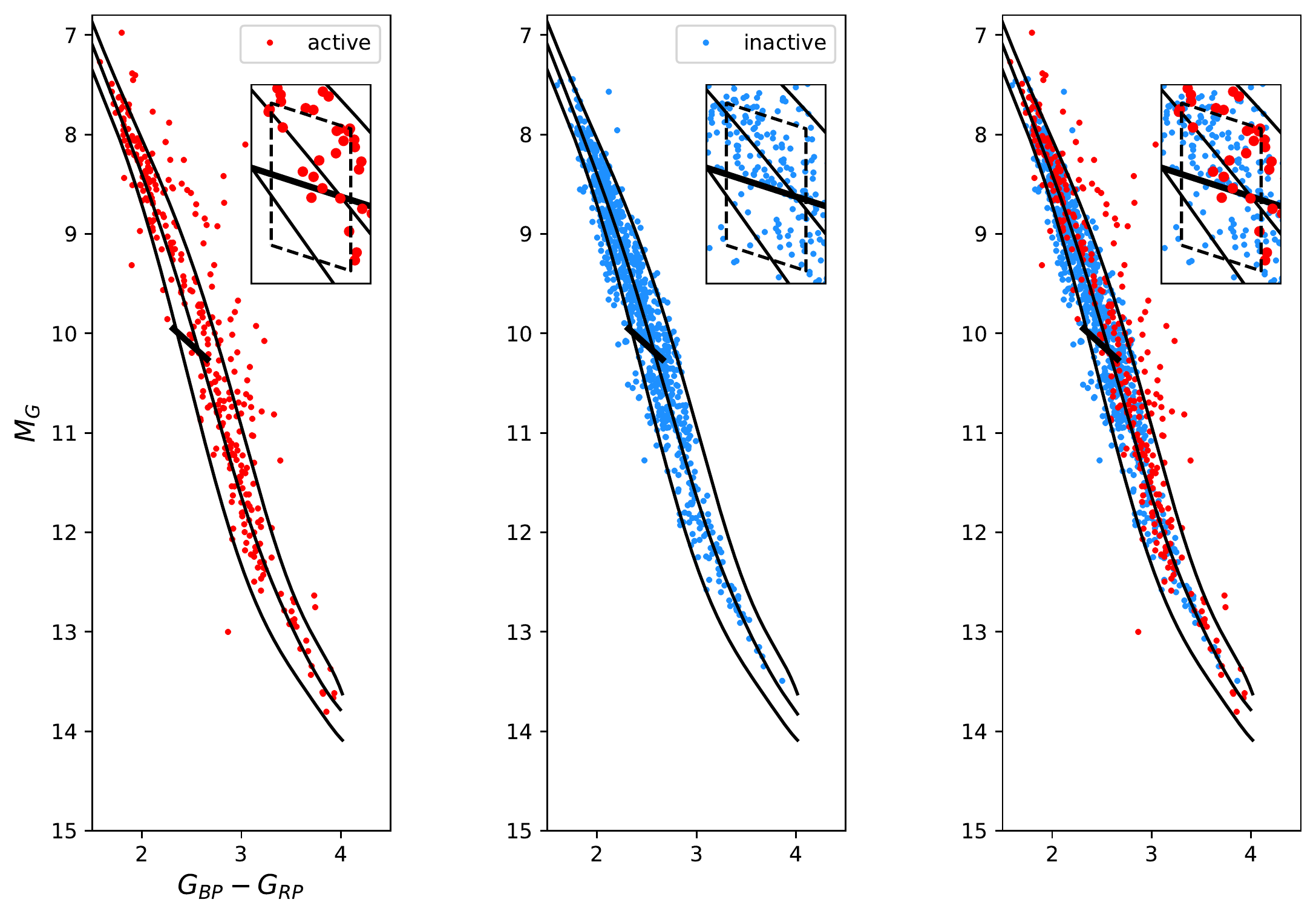}
    \caption{Selected M dwarfs from \citet{Jeffers2018} in the HRD. Symbols and lines are defined in Figure~\ref{fig:ROI} and Figure~\ref{fig:Newton}.}
    \label{fig:CARMENES2}
\end{figure}

The CARMENES survey targets nearby M dwarfs to monitor their long-term radial velocity variations to detect exoplanets. Recently, \citet{Jeffers2018}, hereafter J18, released rotation and activity results of more than 2,200 M dwarfs using the rich CARMENES data. We cross-match their sample against Gaia EDR3 met the following criteria: 1) 1.5$< G_{BP}-G_{RP} <$4, 2) the parallax error is less than 10\% of the parallax, 3) RUWE$<$1.4, and 4) removing known spectroscopic binaries noted in J18. A total of 1,177 stars with a mean parallax of 60.2 mas in EDR3 are shown in Figure~\ref{fig:CARMENES2}. We use the H$\alpha$ EW limit of -0.5\AA~discussed in \citet{Jeffers2018} to separate active and inactive stars, and we identify 333 active stars and 844 inactive stars. A total of 14 active stars from J18 are within our ROI, and three of them are below the transition boundary, which is slightly more than the active stars identified by us and N17. Our nineteen stars are overlapped with J18. Among them, two stars, GAI0505-1200 and GAI1019+1952, are identified as active by both J18 and us. One star, GAI1722+0531, shows H$\alpha$ emission with EW of -0.35\AA~by us but -0.11 by J18. That means this star became active in 2022. All the rest overlapped targets are inactive from both works.

In comparison to results from N17 in Figure~\ref{fig:Newton}, more early active M dwarfs are identified in J18 than in N17, especially in region 1. The ratio of active stars in regions 2 and 1 is 2.87 (95:33) in J18, but the ratio for N17 is infinity (49:0). However, similar to N17, fewer active stars from J18 are in region 3 than in region 4. For stars below the GE, we found that the ratio of inactive stars in regions 4 and 3 is 101:143 (0.71:1), and the ratio of active stars is 81:51 (1.58:1). Hence, these ratios are consistent with the results in N17, where active stars tend to be above the fitted main sequence, especially for stars with $M_{G}$ between 10 and 12.

\subsection{Zhang 2021}
\label{sec:LAMOST}

\subsubsection{Stars in the ROI}
\begin{figure}
    \centering
    \plotone{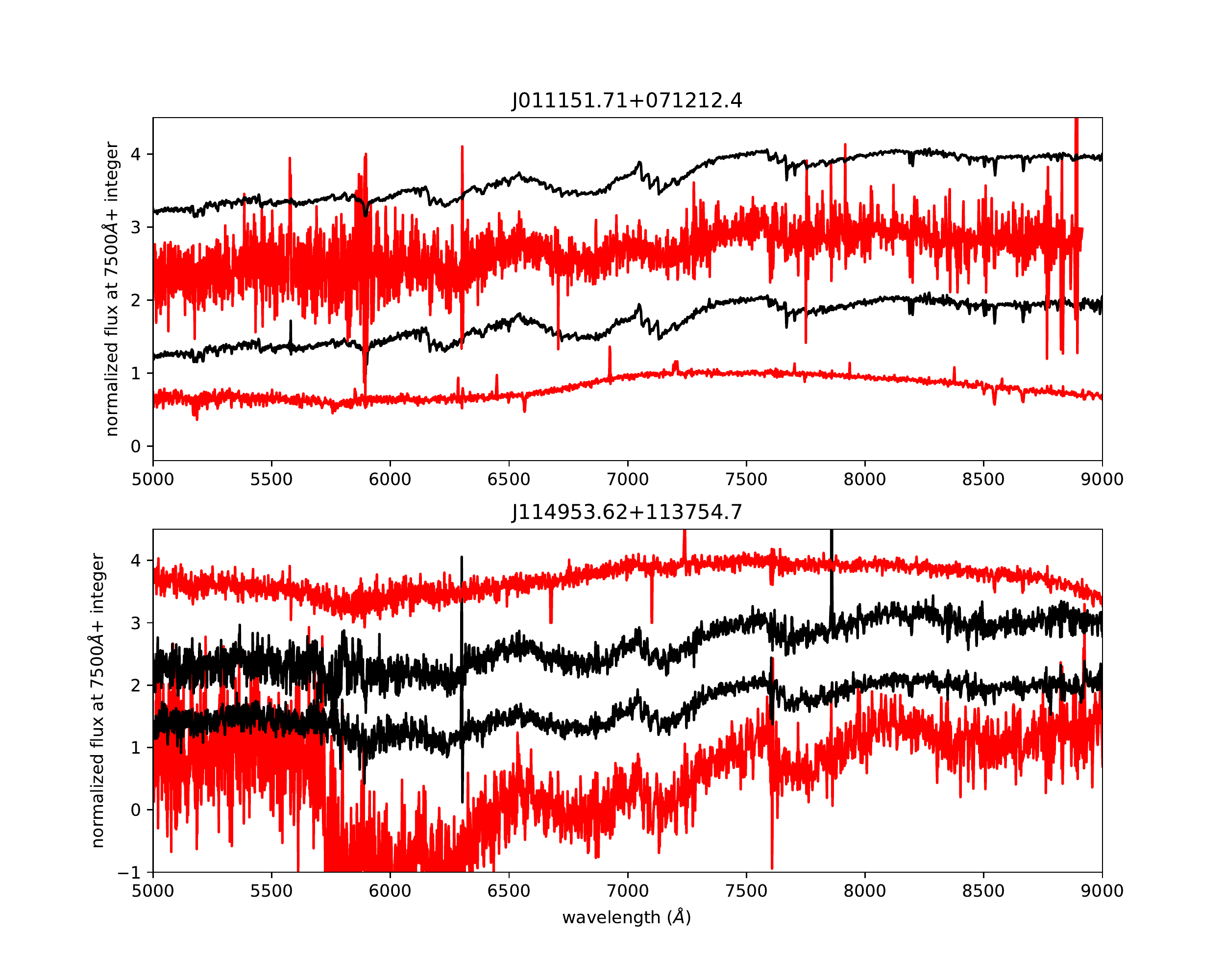}
    \caption{LAMOST spectra of two stars with multi-epoch observations. All spectra are normalized and shifted by an integer. These two stars are classified as M-type in Z21, and obviously, red lines are not M dwarfs. Both targets are inactive.}
    \label{fig:LAMOSTbad}
\end{figure}

\citet{Zhang2021}, hereafter Z21, released chromospheric activities of 738,476 spectral M-type stars, including both dwarfs and giants, based on the LAMOST (Large Sky Area Multi-Object Fibre Spectroscopic Telescope) low and medium resolution spectral surveys. By following the three-step Gaia data quality selecting criteria discussed in \cite{Lindegren2018} and \cite{Jao2018}, we matched 541,124 entries in Z21 with good astrometric data, and a total of 68,141 entries are flagged with H$\alpha$ emissions by Z21. However, we found many stars have multiple entries in their catalog because of multi-epoch observations, and some stars have inconsistent spectral energy distributions. Two example stars are given in Figure~\ref{fig:LAMOSTbad}. Although all spectra are classified as M type, the highlighted spectra appear not as M dwarfs. We think the spectral misclassification could be caused by 1) misplaced fibers, 2) flux and telluric lines calibrations, or 3) mislabeled object identifications. Because their released H$\alpha$ activity results are results of their incorrect spectral classifications, spectra of non-M dwarfs have been mixed in their results. 

Thus, because some stars have been misclassified as M dwarfs in Z21, we re-classify their spectra in order to make a similar H$\alpha$ activity distribution figure like Figure~\ref{fig:Newton} and~\ref{fig:CARMENES2}. We use the same EDR3 selecting criteria discussed in section~\ref{sec:targets} to select stars with high-quality astrometric data. 
We then retrieve spectra from the LAMOST DR6 for limited stars within 100 pc within our ROI as our base sample and use the Sloan Digital Sky Survey (SDSS) M dwarf standard star templates released in \citet{Bochanski2007} to re-classify their spectral types. After re-classifying their spectra, we identify 639 M dwarfs within 100 pc from 751 spectra. Instead of using the H$\alpha$ activity flag reported in Z21 to identify active and inactive stars, we re-identify the H$\alpha$ profile by fitting a skewed Gaussian function between 6,538 and 6,750\AA. The peak of the skewed Gaussian, negative or positive, is an initial indicator of the activity, and we later visually inspect all spectra to distinguish H$\alpha$ emission and absorption. These stars then serve as our reference stars to set the activity limits to separate active and inactive stars for a much larger sample.

In Z21, they measured various parameters for the H$\alpha$ line, including EW (EWHa), S/N (SNHa), and height (Height) of the H$\alpha$ line, as well as the S/N of the spectrum (SNr). Using a combination of these parameters, Z21 set criteria to indicate active and inactive stars among their 738,476 stars. Because the 639 M dwarfs are our base sample, we will redefine the criteria to select active and inactive stars. Without re-measuring these parameters discussed above, we continue to adopt these parameters given in Z21 and present new limits in Figure~\ref{fig:rawlimit}(a), where active stars have 1) ``EWHa'' $>$ 0.58\AA,  2) ``SNHa'' $>$16, 3) ``Height'' $>$ 2.5, 4) SNr $>$7.7, and 5) EWHa $>$ e\_EWHa.  Thus, these criteria are optimized for mid-M dwarfs, and two active stars identified by Z21 do not meet the criteria as seen in the left plot of Figure~\ref{fig:rawlimit}(a). This is because both stars have relatively low SNHa. 

\begin{figure}
    \centering
    \plotone{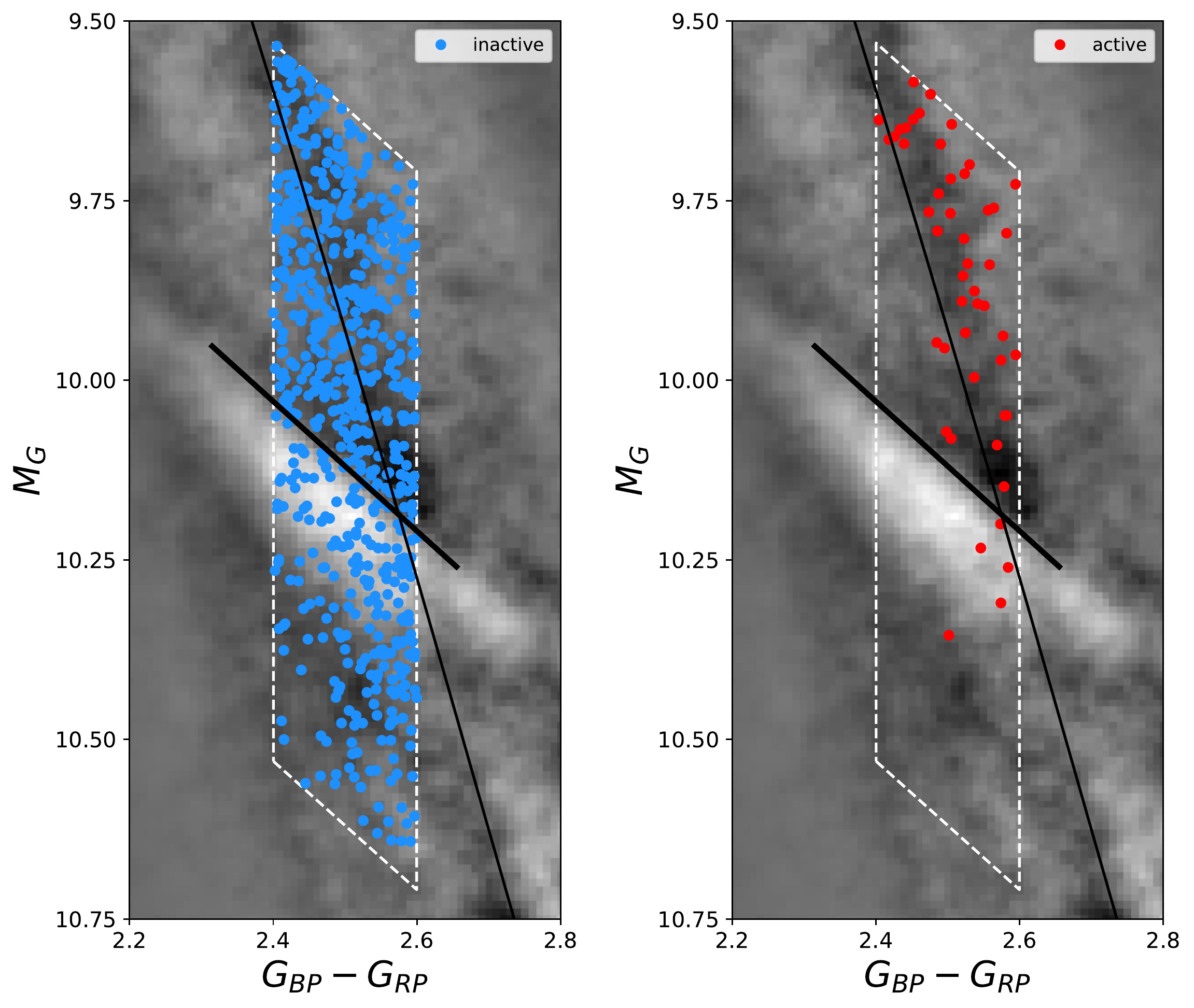}
    \caption{Active and inactive stars in LAMOST are selected using our new criteria. This figure is similar to Figure~\ref{fig:ROI}}
    \label{fig:LAMOSTHa}
\end{figure}

Forty-nine active and 590 inactive stars within the ROI are identified using the new criteria shown in Figure~\ref{fig:LAMOSTHa}. We can see inactive stars spread throughout this ROI. The active stars, however, are generally in the top half of the ROI, especially above the fitted main sequence. Five active stars are in the bottom half of the ROI, much more than the active stars identified by N17, J18, and this work. Nonetheless, the percentage of active stars remains low in the bottom half of the ROI, and there is no activity anomaly in the gap.

There are a total of 67 overlapped stars between Z21 and this work. Four stars, GAI0128+1617, GAI1407+0311, GAI1556+0254, and GAI1722+0531 are identified as active by both projects. Three additional stars (GAI0552-0305, GAI0915+1414, and GAI1452+0629) were also marked as active by Z21, but our spectra indicate they are inactive. We extract spectra for these three stars from LAMOST and found that only GAI0552-0305 is active, which is just above the gap ($M_{G}=10.08$ and $G_{BP}-G_{RP}=$2.5). This star may have become inactive during our observation. All the rest of overlapped targets are identified as inactive by both projects.

\subsubsection{Apply new criteria to the entire M dwarfs on the main sequence}

\begin{figure}
    \centering
    \gridline{\fig{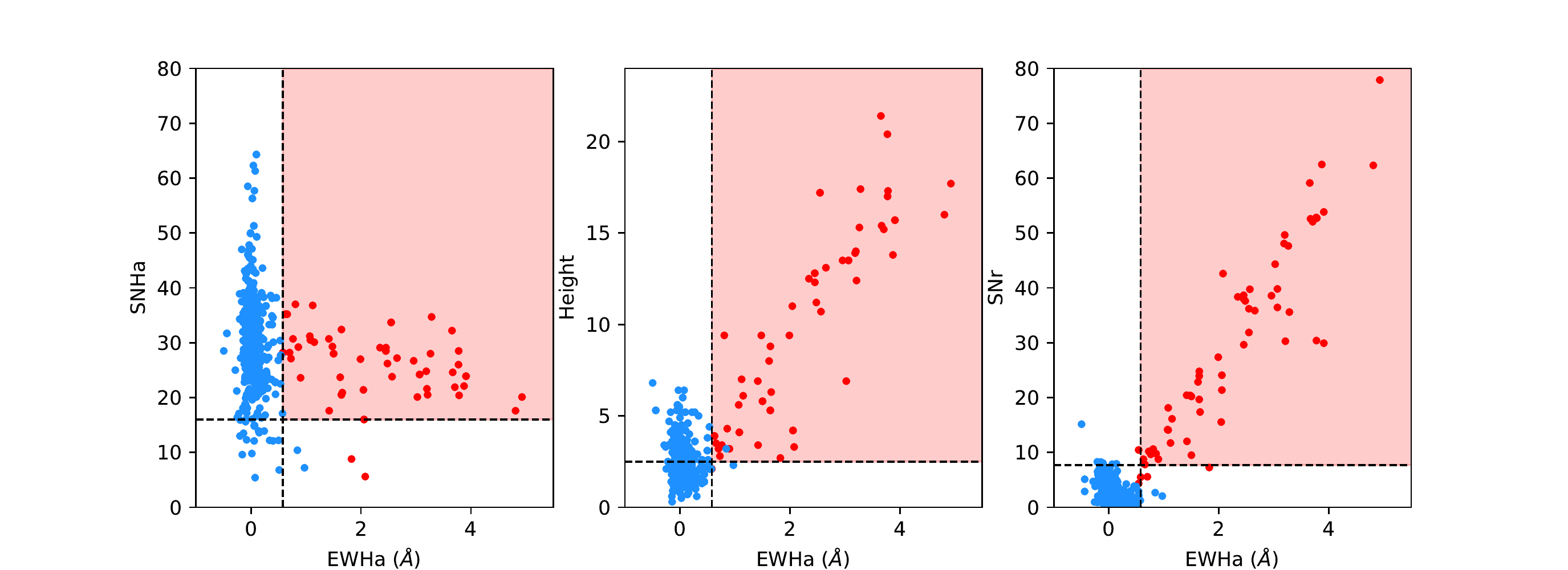}{0.9\textwidth}{(a) New selection criteria}}
    \gridline{\fig{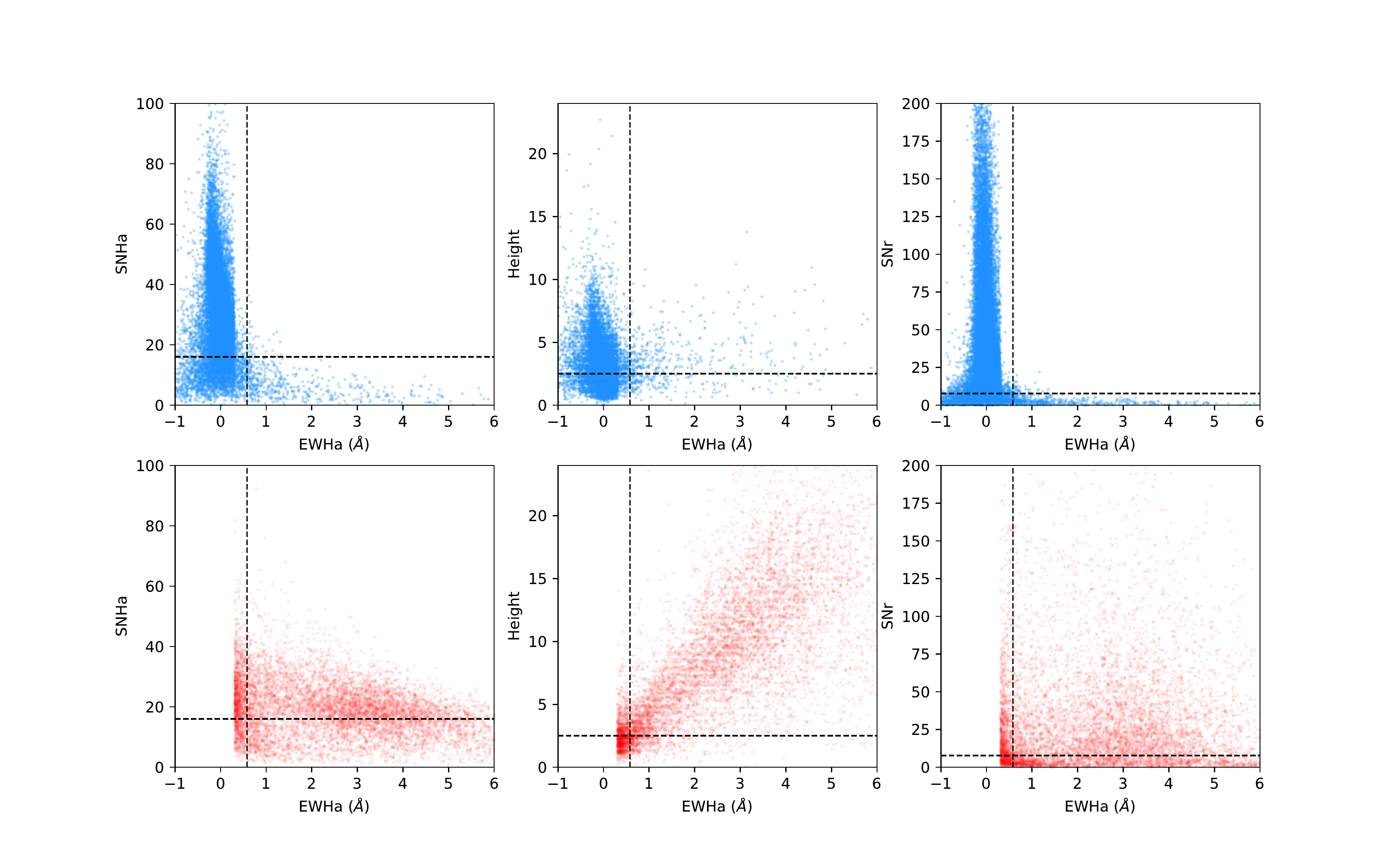}{0.9\textwidth}{(b) Select potential active stars}}    
    \caption{Active and inactive stars in different H$\alpha$ parameter spaces, i.e, EWHa, SNHa, Height, and SNr. (a) New criteria based on re-classified 639 M dwarfs within 100 pc. New limits are marked in black dashed lines. The blue and red dots represent inactive and active stars determined by Z21, respectively. The colored regions are defined as active stars using the new criteria defined in this work. (b) The top panel is for inactive stars (blue dots), and the bottom panel is for active stars (red dots) originally flagged in Z21. The dashed lines mark the new criteria given in (a). This shows some of the active stars flagged by Z21 won't be identified as active by our new criteria.}
    \label{fig:rawlimit}
\end{figure}

\begin{figure}
    \centering
    \plotone{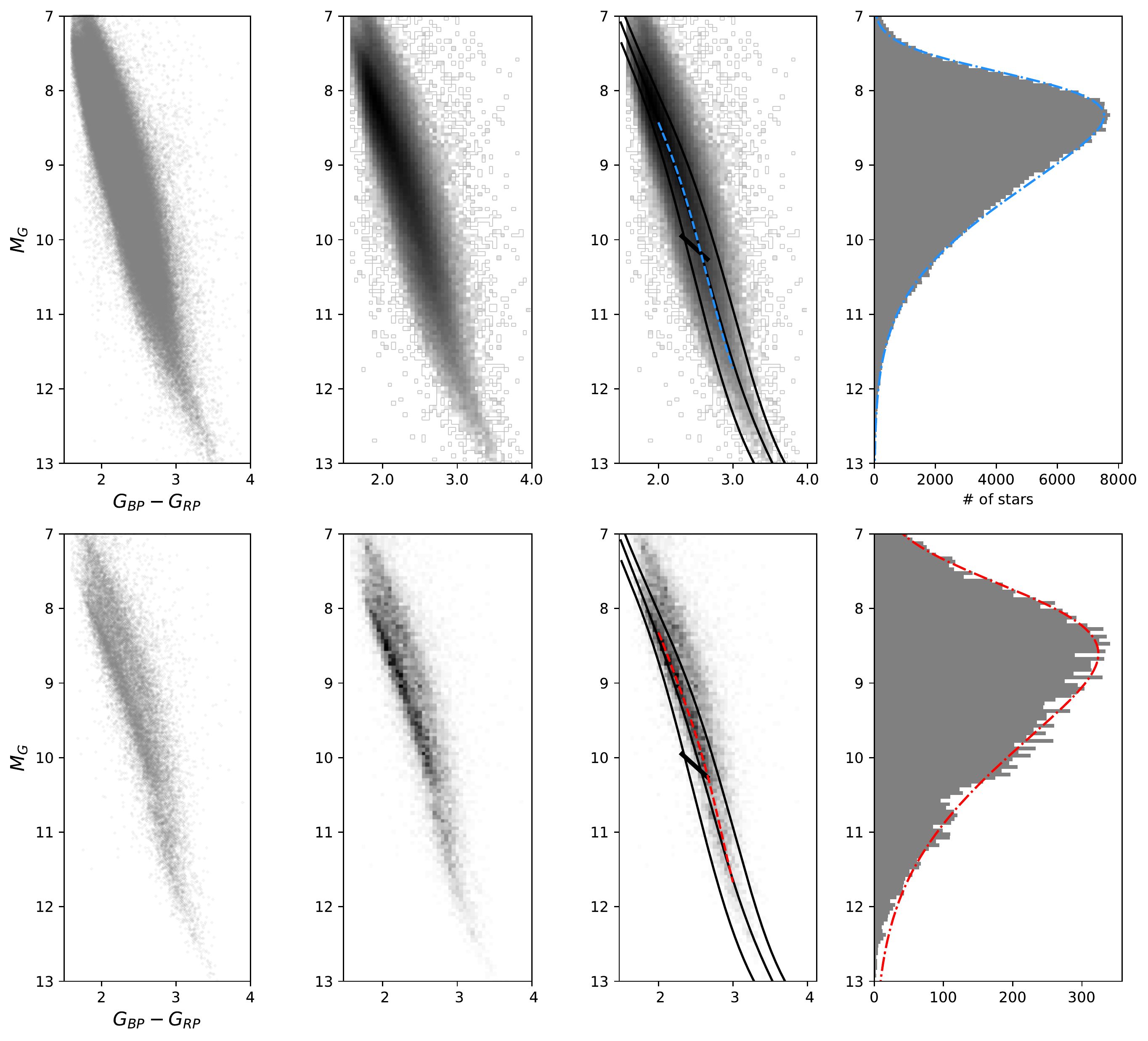}
    \caption{Distributions of inactive (top panel) and active (bottom panel) stars in the HRD from the LAMOST using our new criteria. The two left plots are scattering plots for inactive and active stars. The center four plots are two-dimensional histograms with a bin size of $\sim$0.05 magnitude at $M_{G}$ and $G_{BP}-G_{RP}$. The two-dimensional histograms for active stars are on a linear scale but on a log scale for inactive stars. Four black lines are the same lines defined in Figure~\ref{fig:Newton} are shown in two of the center plots, and they are served as a guide for the distribution of the main sequence and the gap. The red and blue dashed lines represent the best-fitted distribution for active and inactive stars, respectively. The right two plots are the histograms of $M_{G}$ with a bin size of 0.05 mag. Two dash-dotted lines represent the fitted distribution for inactive and active stars.}
    \label{fig:Zhangtight}
\end{figure}

Figure~\ref{fig:rawlimit}(b) compares activity criteria defined by Z21 and new criteria by us in different parameter spaces. Although Z21 noted that stars with EWHa greater than 0.75\AA~ are selected as active stars, this figure shows the limit they used is $\sim$0.3\AA. By applying the new criteria to entire dwarfs in Z21 without re-classifying the spectra types and re-measuring H$\alpha$ parameters of 333,044 stars with $G_{BP}-G_{RP} >$ 1.6 and $M_{G}>$6.0 in Gaia EDR3, we determine 18,472 stars are active, and the remaining are inactive. Their distributions are shown in Figure~\ref{fig:Zhangtight}, where the top panels are inactive stars, and the bottom panels are active stars. 

Ninety-five percent of stars are inactive, distributed throughout the main sequence without showing any specific pattern. The fitted distribution of inactive stars shown as a blue dashed line generally follows the fitted main sequence, so the distribution or populations of inactive stars represents the main sequence. In contrast to inactive stars in Figure~\ref{fig:Zhangtight}, the distributions of active stars, which is 5\% of the sample, in the HRD are mainly composed of two different bands where one band is the dwarfs on the main sequence, and the other is possibly equal mass unresolved binaries elevated above the main sequence. Active stars above the gap are mostly in region 2 (defined in Figure~\ref{fig:Newton}), and the density contrast between regions 1 and 2 is high. This indicates active stars above the gap are mostly elevated above the fitted main sequence. The best-fitted distribution for active stars is a red dashed line in the bottom panel. Unlike the best-fitted line for inactive stars, the fitted line for active stars does not match the main sequence and is shifted 0.07 mag redder in $G_{BP}-G_{RP}$ color at about $M_{G}\sim$10. For active stars below the gap, we can see the distribution is gradually curved to match the distribution of the main sequence. In general, the distributions of active and inactive stars from Z21 in the HRD are consistent with the distributions based on results from N17 and J18 shown in Figure~\ref{fig:Newton} and~\ref{fig:CARMENES2}.

The two rightmost plots in Figure~\ref{fig:Zhangtight} are histograms for inactive and active stars. The two dash-dotted lines represent their distributions fitted by using a skewed Gaussian function. We can see that the distribution of inactive stars is relatively smooth, but surprisingly, there seems to be a prominent dip at $M_{G}\sim$10.5 in the histogram for active stars. We will discuss more of this newly identified activity dip in section~\ref{sec:newfeature}. 

\subsection{Reiners 2022}

\begin{figure}
    \centering
    \plotone{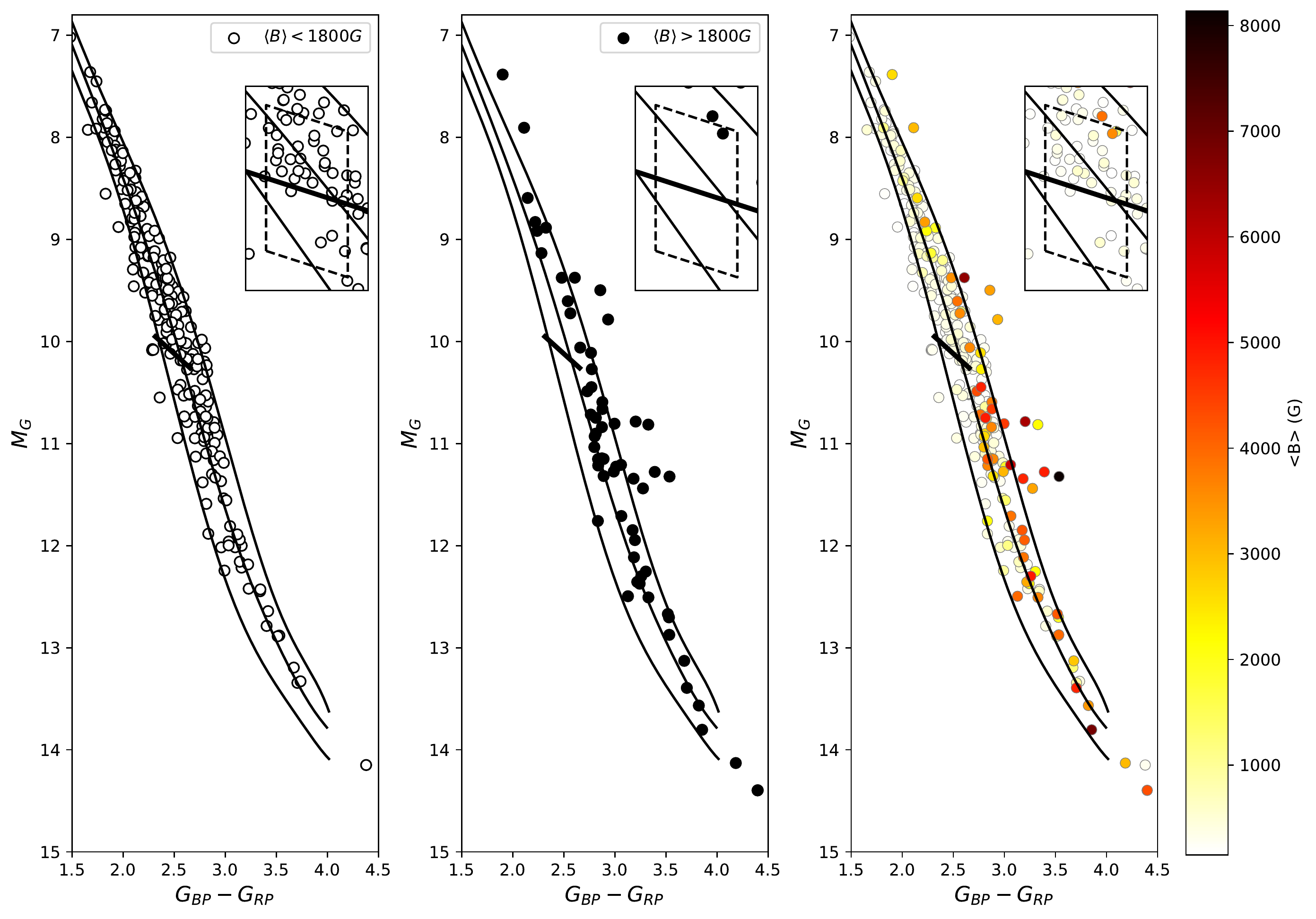}
    \caption{M dwarfs from \cite{Reiners2022} in the HRD. The left plot shows stars with $\langle B \rangle<$ 1800 G in open circles, and the center plot is for stars with $\langle B \rangle>$ 1800 G in filled circles. Because of the relation between magnetic fields and H$\alpha$ fluxes, the distributions shown here are proxies of H$\alpha$ lines. The right plot shows the entire samples in \cite{Reiners2022} and the strengths of magnetic fields in colors. The black lines and ROI are defined in Figure~\ref{fig:ROI} and Figure~\ref{fig:Newton}.}
    \label{fig:Reiners2022}
\end{figure}

\cite{Reiners2022}, hereafter R22, reported mean-surface magnetic fields of 292 nearby M dwarfs in the CARMENES project and an additional 22 stars from the literature. The mean parallax of this sample is 98.5 mas, so their work comprises the nearest M dwarfs among all works discussed in this manuscript.  They found a linear power law relation between the magnetic fields and $L_{H\alpha}$ in their Figure 9, and a magnetic field of $\sim$1800 G could separate fast and slow rotators. Hence, using this magnetic field limit, we could separate their sample into two populations: one contains fast-rotating stars with relatively stronger magnetic fields and high H$\alpha$ luminosities, and the other group has relatively weaker magnetic fields and H$\alpha$ luminosities. We then plot their samples using this limit to separate these two populations in the HRD in Figure~\ref{fig:Reiners2022}.

The left plot of Figure~\ref{fig:Reiners2022} shows they have 27 slow rotators with $\langle B \rangle <$ 1800 G within the ROI, and they have only one faster rotator with $\langle B \rangle >$ 1800 G within the ROI. Unlike the three other works discussed in this section, the relatively small sample size within the ROI from R22 is hard for us to study and compare activity within our ROI. Nonetheless, by examining the activity distribution across the main sequence, we find slow rotators with $\langle B \rangle <$ 1800 G are generally distributed on either side of the main sequence, but fast rotators with $\langle B \rangle >$ 1800 G are elevated in the top half of the best fitted main sequence or region 2 and 4, with a few exceptions at the lower main sequence. 

In summary, all four earlier works from N17, J18, Z21, and R22 show active partially convective stars are mostly elevated above the main sequence, but active fully convective stars are on either side of the main sequence. Other than the samples in Z21, the other three works don't have enough stars within our ROI. However, all these works consistently show active stars are mostly in the top half of the ROI, and the bottom half of the ROI, including stars in the gap, lacks active stars.

\section{A dip in the H-alpha activity distribution}
\label{sec:newfeature}

\subsection{confirmation of the activity dip}

This activity dip in Figure~\ref{fig:Zhangtight} is seen using our revised criteria to define active and inactive stars. We found that when we use the active and inactive stars originally flagged by Z21 (See Appendix~\ref{sec:appendex1}), both old and revised selection criteria yield the same dip at the same location. In spite of that, we understand a sampling bias could also create an expected distribution dip. For example, \cite{West2011} studied a large number of M dwarfs using data from SDSS, which has similar observational approaches and goals as the LAMOST, and reported a selection bias in their SDSS sample so that a population gap is seen on the main sequence. However, the smooth distribution of inactive stars shown in Figure~\ref{fig:Zhangtight} likely eliminates sampling biases in LAMOST.

This activity dip mainly comes from stars within 200 pc in Z21. For stars beyond 200 pc or stars fainter than $M_G\sim$10.6, the number of stars drops rapidly because of the magnitude limit from LAMOST. To further confirm this activity dip, we select two equal-volume samples to compare: distances of 0--158.7 and 158.7--200 pc. We found this activity dip is seen at the same location regardless of the sample. Besides, LAMOST data also include stars in nearby young clusters, i.e., Pleiades, M44, and M67. After stars in those parts of the sky are excluded, the activity dip still exists at the same location on the main sequence.

Finally, the dip in the histogram can also be identified in $M_{BP}$ and $M_{RP}$ at a range of 12.0--12.25 and 9.0 -- 9.55, respectively. The dip is the most prominent in $M_{RP}$ because Gaia's RP filter has the highest transmission at H$\alpha$ among all three Gaia filters. Consequently, we can reproduce this activity dip using various samples. We note that this activity dip can't be identified in N17 and 18, and we suspect these two works have relatively smaller sample sizes to reveal this feature.

\subsection{In-depth analysis of the activity dip}

\begin{figure}
    \centering
    \includegraphics[scale=0.55]{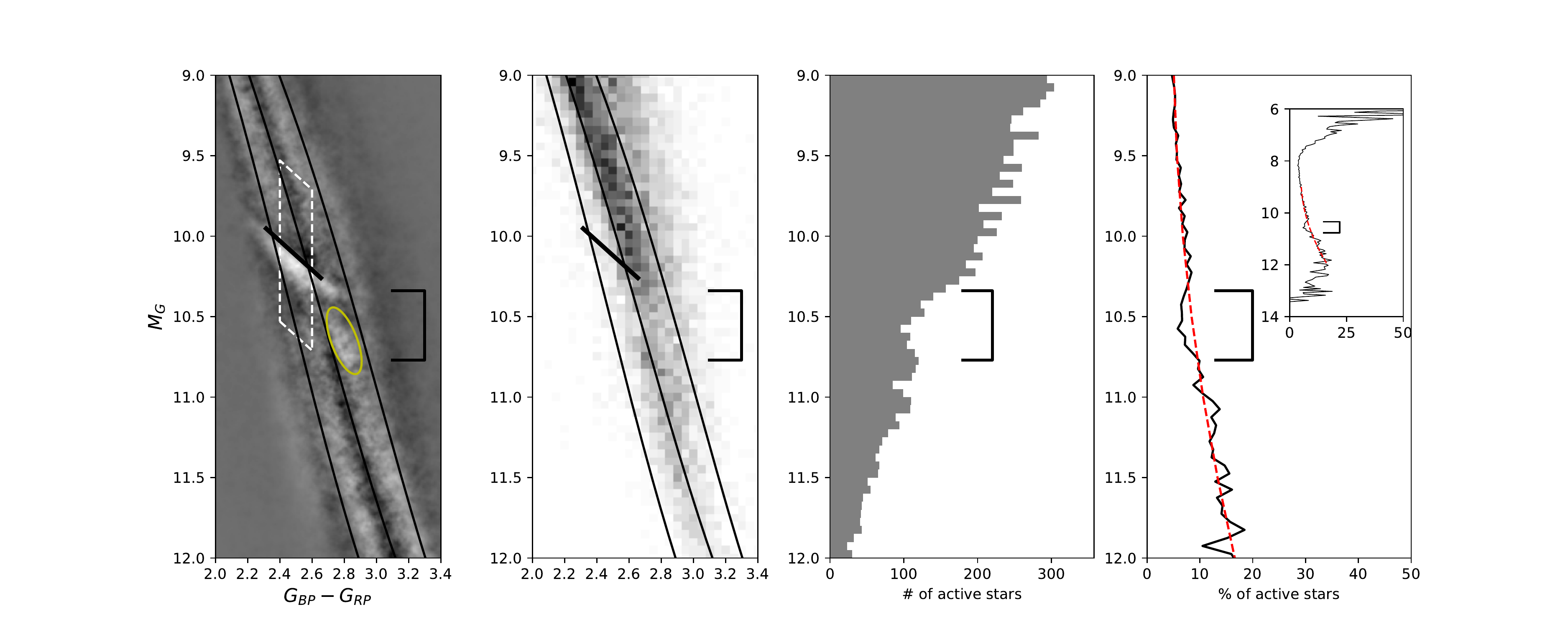}
    \caption{(Left): The activity dip is marked as a square bracket on the enhanced HRD. This bracket is shown in all four plots. A stellar low-density region first identified in \cite{Jao2020, Jao2021} is marked in yellow. The dashed box is our ROI. (Center Left): A portion of the two-dimensional histogram of active stars in the bottom panel of Figure~\ref{fig:Zhangtight}. (Center Right): A histogram of active stars. Both center left and center right plots are the same as the right two plots in the bottom row of Figure~\ref{fig:Zhangtight}. (Right): A distribution of percentages of active stars. The inset plot covers a larger range of this distribution. Red dashed lines are the second-order polynomial fits of the percentages vs $M_{G}$ from 9 to 12 magnitude.}
    \label{fig:newfeature}
\end{figure}

Figure~\ref{fig:newfeature} illustrates the location of the activity dip on the HRD. In the left plot of this figure, this activity dip appears to overlap a low-density region on the main sequence first identified by \cite{Jao2020} and \cite{Jao2021}. This low-density region is not as prominent as the gap but is statistically significant. The cause of this low-density region is yet to be known, and we have no prior reason to relate this H$\alpha$ activity dip to this low-density region. Here, we only highlight the coincidence of these two features on the lower main sequence.

Other than studying the number of active stars in each bin, we show the percentage distribution of active stars in Z21 in the inset in the right plot of Figure~\ref{fig:newfeature}. The percentage distribution peaks at $M_{G}\sim$6, and decreases till $M_{G}\sim$8, where it corresponds to the M0V. Then, the percentage rises until about $M_{G}\sim$12 and decreases again till the end of the main sequence. The fall and rise of the percentage of active stars could be related to the chromospheric activities, the line strengths between H$\alpha$ and the continuum, and the LAMOST survey's incompleteness toward the end of the main sequence. Nonetheless, there is a prominent activity dip between $M_{G}=$ 10.3 and 10.8, and we estimate 3\% of fewer stars are active compared to the best-fitted distribution at $M_G\sim$10.6 or at the deepest dip.

The activity dip marked as square brackets in all four plots in Figure~\ref{fig:newfeature} is just below the gap, and this region overlaps a large portion of the bottom half of our ROI. Consequently, our almost null H$\alpha$ emission detection in the bottom half of the ROI is consistent with the activity dip in Figure~\ref{fig:newfeature}. In other words, stars in this activity dip region are fully convective M dwarfs. When Z21 used spectral sub-types to study the percentages of active M dwarfs on the main sequence, they reported a similar drop at M4 (see their Figure 4), but they thought this was due to a change in the magnetic dynamo mechanism at the partially and fully convective boundary. However, based on the markings in Figure~\ref{fig:newfeature}, this region is lower than the gap or boundary.

Thus, putting this all together, we find that the most luminous fully convective M dwarfs, those with $M_{G}$ between 10.3 and 10.8 appear less active than stars above and below this $M_{G}$ range. In other words, some most massive fully convective M dwarfs have their H$\alpha$ activities ceased, or no emission can be detected by the resolution and observation limits of CHIRON and LAMOST.

\section{Rotation and H$\alpha$ Emission Relative to the Activity Dip}

Stellar activity and rotation or flares are known to have a broken power law relation \citep{Reiners2014, Newton2017, Wright2018, Reinhold2020, Namekata2022}. To further understand their relations, we select two recent works\footnote{\cite{Popinchalk2021} also assembled a large number of rotation periods for young and field stars. The reason for excluding it in finding the cause of this dip is fully discussed in Appendix~\ref{app:Popinchalk}.} from \cite{Lu2022}, and \cite{Reinhold2020} that contain a large number of rotation periods from two different projects.

\subsection{Lu et al.~(2022) --- Linking Activity and Rotation}

\begin{figure}
    \centering
    \includegraphics[scale=0.7]{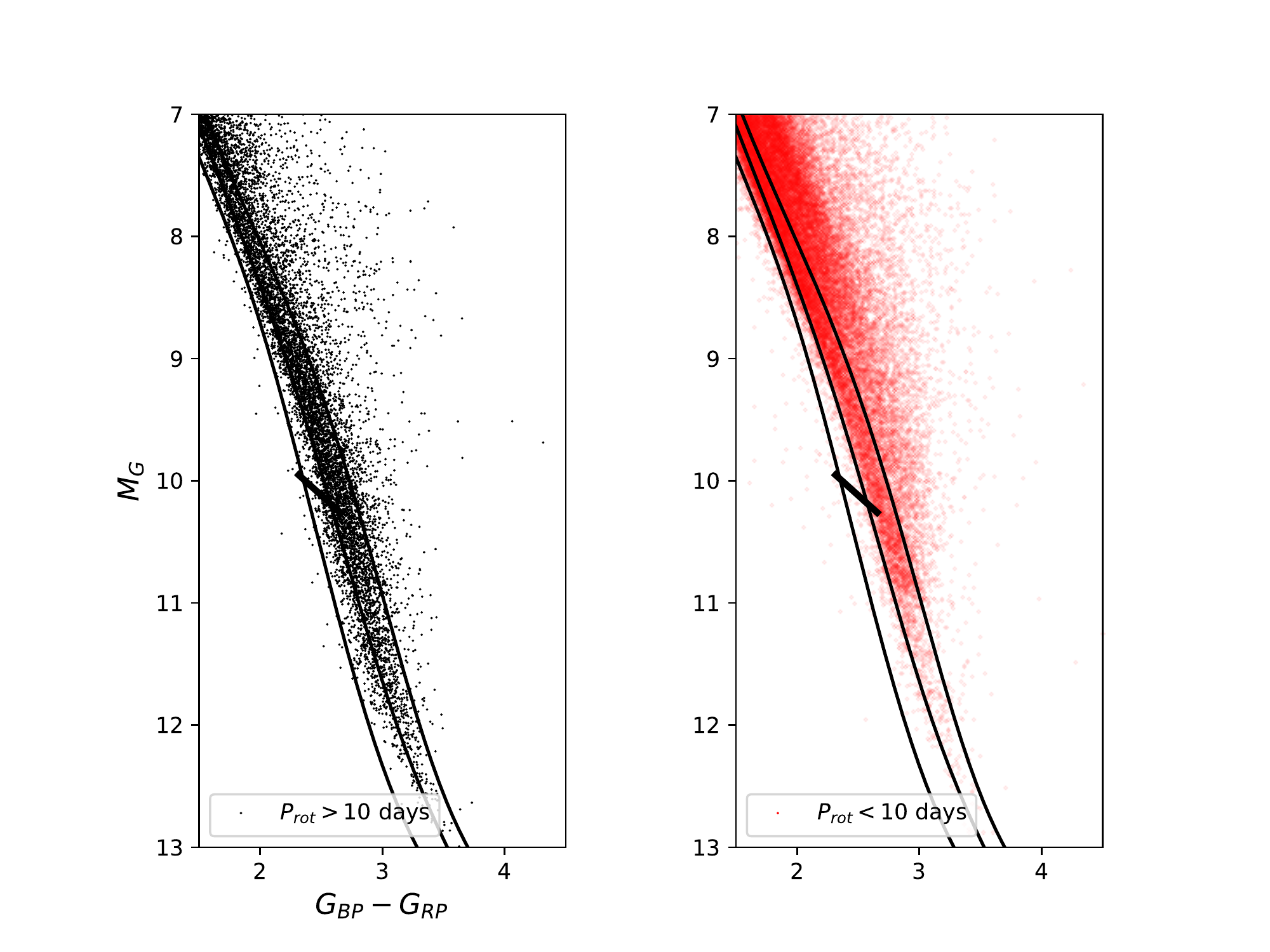}
    \caption{Stars from \cite{Lu2022} on the HRD. The left and right plots are for stars with $P_{rot}>$ and $P_{rot}<$ 10 days, respectively. \cite{Lu2022} showed that 5\% of the periods longer than 10 days and 50\% of the periods shorter than 10 days are possibly incorrect.}
    \label{fig:ZTF}
\end{figure}

\cite{Lu2022}, hereafter L22, reported rotation periods for over 40,553 stars measured using the Zwicky Transient Facility (ZTF). Two distributions of their targets on the HRD after matching Gaia DR3 are shown in Figure~\ref{fig:ZTF}, and the mean parallax is 3.8 mas. In this figure, we split their sample into two groups, $P_{rot}>10$ and $<10$ days. We can see that both groups are obviously elevated above the best-fitted main sequence in regions 2 and 4, and relatively fewer stars are in regions 1 and 3. Apparently, this distribution is strikingly similar to those active stars in N17, J18, and Z21 and stars with $\langle B \rangle >$ 1800 G in R22. ZTF is a ground-based observation, so large and prominent spot modulations could preferably be detected. Hence, those spot modulations detected by ZTF are closely relative to active stars, and we would assume most of these stars could be considered active. 

L22 stated that 50\% of the rotation periods less than 10 days are possibly incorrect, and 5\% of the periods longer than 10 days are erroneously measured in their work, so it is hard for us to take advantage of this rich dataset to investigate this activity dip further. Furthermore, we don't have enough relatively fast rotators or potential inactive stars to compare with. In spite of these challenges, the distribution of stars in L22 indicates a strong connection between rotation and stellar activity on the HRD.

\subsection{Reinhold \& Hekker~(2020) --- Linking H$\alpha$ activity to Fast Rotators}

\begin{figure}
    \centering
    \includegraphics[scale=0.7]{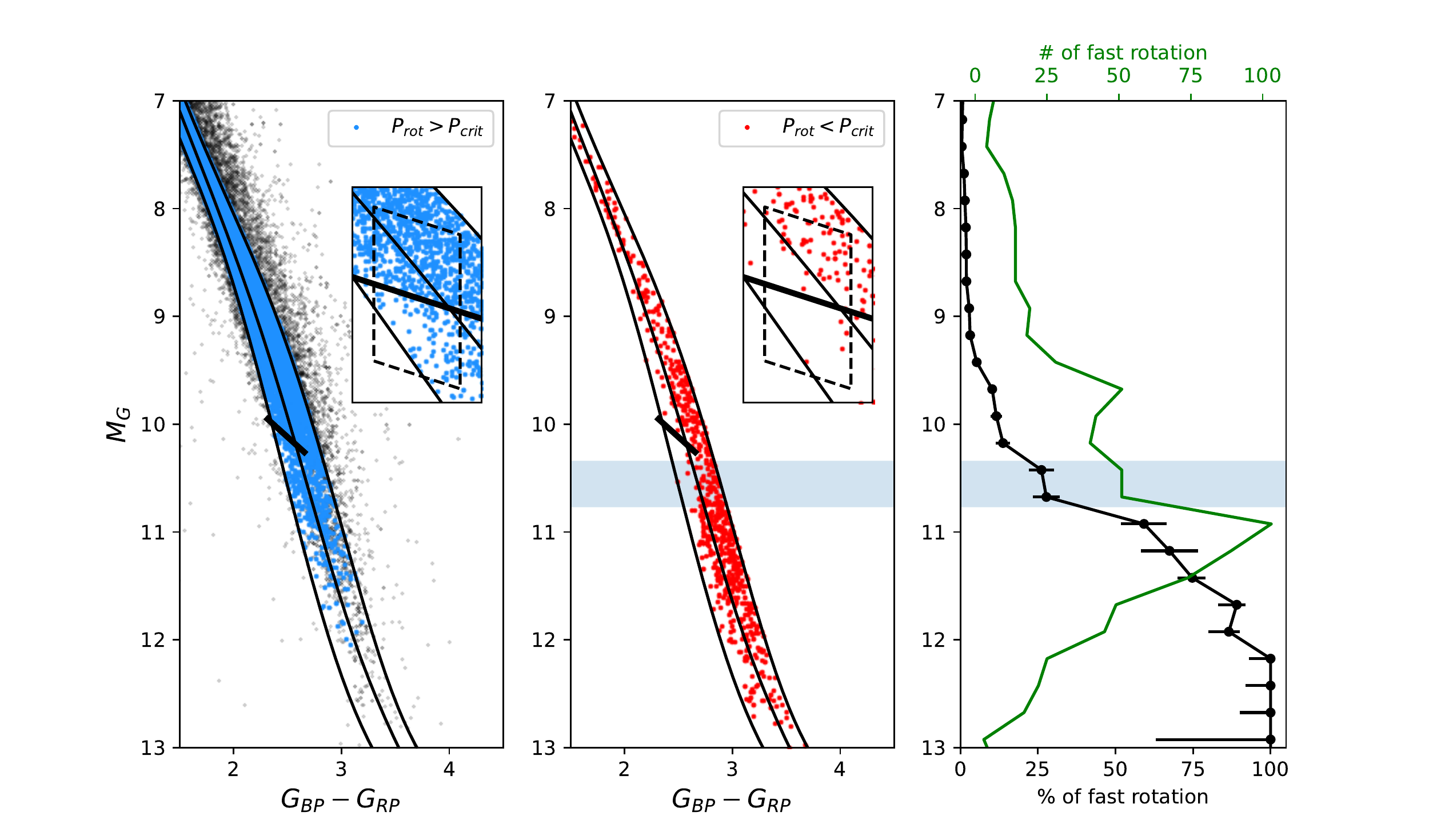}
    \caption{The entire sample in \cite{Reinhold2020} is shown in the left plot. Because their targets include various young and old dwarfs selected in different K2 Campaigns, we only focus on stars falling between the upper and lower envelopes of the main sequence. Stars excluded from our statistics are plotted in black dots. The two left plots are for stars with periods greater (blue dots) or less (red dots) than the critical period ($P_{crit}$). The right plot shows the percentage of fast rotation stars in each bin of $\Delta M_G$=0.25 in a black line. The error bars are calculated using the same method discussed in Table~\ref{tbl:percentage}. The blue bands mark the H$\alpha$ activity dip in LAMOST. The green line is the number of fast rotators in each bin. Inserted plots highlight stars within the ROI in this work.}
    \label{fig:Reinhold}
\end{figure}

\cite{Reinhold2020}, hereafter RH20, reported rotation periods of more than 29,860 stars from the $Kepler$ K2 Campaigns from 0 to 18 with a mean parallax of 5.1 mas from Gaia DR3, so results from their large numbers of M dwarfs provide another avenue to understand H$\alpha$ activity distribution on the main sequence. Because of the experimental design of the K2 mission, all targets are mainly in the ecliptic plane and include a few young clusters like Pleiades, M44, M67, and small portions of Hyades and Taurus-Auriga. We only consider stars with RUWE$<$1.4 that fall between the upper and lower envelopes of the main sequence to represent the overall population of the main sequence stars. Figure~\ref{fig:Reinhold} shows these selected 24,836 targets with $M_G>$ 7 on the HRD. Unlike stars in L22, targets with long periods in RH20 are distributed on either side of the best-fitted main sequence. Therefore, we can assume stars in RH20 potentially contain both active and inactive stars, but L22 mostly contains active stars.

\subsubsection{Split stars in RH20 into fast and slow rotators}

The broken power law relations between rotation and activity are fully discussed and presented in \cite{Reiners2014} and \cite{Newton2017}, using X-ray and H$\alpha$ fluxes, respectively. The other difference between these two works is whether the Rossby number is used to represent this relation or not. Nonetheless, the X-ray or H$\alpha$ activity is saturated when the Rossby number or rotational period is less than a critical value. As discussed in \cite{Jao2022}, we face challenges in determining the convective turnover time, which is used to obtain the Rossby number. Furthermore, \cite{Douglas2014} showed the broken power laws of X-ray and H$\alpha$ activities for stars in young clusters have almost the same Rossby number or critical value at the kink. Hence, instead of using the critical Rossby number reported for the H$\alpha$ line in \cite{Newton2017}, we adopt the critical value reported in \cite{Reiners2014} based on X-ray activities so that we can avoid the difficulties of obtaining convective turnover time and eventually the Rossby number.

The breakpoint or the critical value is at $\log(L_X/L_{bol})$=$\log(k R^{\alpha}P^{\beta})$=$-$3.14 \citep{Reiners2014}, where R is the radius of a star in solar radius, P is the period in days, $k=1.86\times10^{-3}$, $\alpha=-$4, and $\beta=-$2. Apparently, it shows although the breakpoint is a fixed value for the activity-rotation relation, it also depends on stellar radius and period. As a result, for a given star, we would need to estimate its critical rotational period, $P_{crit}$, so that we can compare its measured rotational period to $P_{crit}$ to understand how active a star is. This also means the $P_{crit}$ is not a fixed value but depends on stellar radii. Without knowing their true H$\alpha$ EWs or fluxes, this method can only distinguish potential saturated and un-saturated X-ray or H$\alpha$ activities. Also, because all the H$\alpha$ activity studies separate targets into two groups, active and inactive, using the $P_{crit}$, we could also help us to split stars in RH20 into two groups of fast and slow rotators. 

We calculate stellar radii for stars in RH20 based on the empirical $M_{G}$ and radius relation in \cite{Pecaut2013}. We then can obtain the $P_{crit}$ for a given target based on the broken power law relation in \cite{Reiners2014}. Stars with measured rotational periods greater than $P_{crit}$ are shown in blue dots on the left of Figure~\ref{fig:Reinhold}. Stars rotate faster than the $P_{crit}$ are shown in red in the middle plot of Figure~\ref{fig:Reinhold}. 

We can see that slow-rotation stars ($P_{rot}>P_{crit}$) are distributed evenly on either side of the best-fitted main sequence or in all four regions. The number of slow rotators decreases toward the lower main sequence. In contrast, fast-rotation stars are mostly elevated above the main sequence for stars above the gap in region 2, but fast-rotation stars below the gap can be seen on either side of the main sequence in region 3 and 4, especially toward the end of the main sequence. Consequently, the distributions of slow and fast-rotation stars on the main sequence are strikingly similar to the distribution of H$\alpha$ inactive and active stars shown in N17, J18, and Z21. The distribution of the fast rotators in RH20 is also similar to the left plot in Figure~\ref{fig:ZTF}.

\subsubsection{Connect H$\alpha$ activity and stellar rotation}

For stars in the ROI, the distributions of slow and fast-rotation stars in Figure~\ref{fig:Reinhold} also resemble the H$\alpha$ activity distribution we found in Figure~\ref{fig:results}. Slow rotators and H$\alpha$ inactive stars above the GE are distributed on either side of the ROI, but the fast rotators and H$\alpha$ active stars are distributed mainly in the top half of the ROI and above the best-fitted main sequence in region 2. No rotation anomaly is seen for stars in the gap or zone E. Consequently, rotation and H$\alpha$ activities are closely related not only in a mathematical relation but in their distributions on the HRD.

The right plot in Figure~\ref{fig:Reinhold} shows the percentage of fast rotators as a black line in a bin of $\Delta M_G=$0.25. This curve shows the percentages of fast rotators are relatively lower in the range of $M_{G}=$ 10.3 and 10.8 than those stars in the bin of $M_G\sim$10.9. This implies fewer stars in this range rotate faster than the $P_{crit}$, which is about 13 days if $M_G=$10.5 is used. The location is surprisingly identical to the H$\alpha$ activity dip highlighted in blue boxes. More importantly, this fast-rotator deficient region is outside of the main sequence gap, so the cause of the deficiency is not due to interior instability.

Furthermore, targets from Z21 and RH20 have no overlap. Targets in Z21 are all over the northern sky, but targets in RH20 are mainly in the ecliptic plane. Yet, the dips identified from these two datasets are at the same location on the main sequence. This further demonstrates that stellar rotation and H$\alpha$ activities are strongly related. A slow or no rotation star produces a weaker H$\alpha$ line or absorption in the atmosphere, and a fast rotator generates stronger dynamos to potentially saturate H$\alpha$ line. \cite{Pass2022} used wide binaries to identify ages of M dwarfs with rotation periods and found M dwarfs with masses between 0.2 and 0.3M$_{\odot}$ spin down in roughly three stages: $P_{rot}<$2 days at 600 Myrs, 2$<P_{rot}<$10 days at 1--3 Gyrs, and rapidly spin down at older ages. The H$\alpha$ activity dip and low percentage of faster rotators in the range of $M_{G}=$ 10.3 and 10.8, where these stars are fully convective stars, are close to the estimated upper mass limit in \cite{Pass2022}. Therefore, some of the M dwarfs with ages older than 3 Gyrs in this magnitude range have spun down rapidly, so fewer stars are active and fast rotators.

We understand the percentage distribution of active stars in Figure~\ref{fig:Zhangtight} is different from the percentage distribution of fast rotators in Figure~\ref{fig:Reinhold}. This is mainly because 1) the split of fast and slow rotators in RH20 can only help us identify potential saturated or un-saturated H$\alpha$ fluxes, and 2) RH20 has incompleteness of slow rotators, especially in the lower main sequence. Hence, the percentage of active stars in Figure~\ref{fig:Zhangtight} rises the most at $\sim$20\%, but the percentage of fast rotators in Figure~\ref{fig:Reinhold} can rise up to 100\%. However, when we focus on the numbers of fast rotators between $M_G=$ 9.5 and 11, we still have fewer fast rotators in this region, and the location is overlapped with the activity dip in Z21. Based on the work in L22 and RH20, we propose the H$\alpha$ activity dip results from lacking fast rotators in this mass range.

\section{The possible cause of the activity dip}

The leading model to explain the stalled or reduced spin down for partially convective stars with masses between 0.4 $M_{\odot}$ and 0.8 $M_{\odot}$ is the transfer of angular momentum from the radiative core to the convective envelope to compensate for the angular momentum loss due to wind braking \citep{Spada2020}. In other words, wind braking develops differential rotation or velocity gradient, where the angular velocity of the radiative zone is greater than the angular velocity of the convective zone. Observationally, this model has generally accounted for the rotation stalling of some stars in nearby clusters \citep{Curtis2020, Dungee2022}, but this two-zone model is designed for stars that are only partially convective. However, the concept of transferring angular momentum from the core, or the lack of it, could be applied to fully convective stars. As discussed in \cite{Dungee2022}, due to the disappearance of the radiative core for fully convective stars, they may lose the reservoirs of angular momentum in the radiative core. Consequently, the wind braking could expedite the spin down for these early types of fully convective stars. The more stars spin down fast, the fewer H$\alpha$ emissions can be detected in the result of the activity dip we see in 
Figure~\ref{fig:Zhangtight}. According to the $^3He$ instability model \citep{Baraffe2018}, M dwarfs with masses of 0.3 -- 0.32$M_{\odot}$ could briefly form a radiative zone approximately between the ages of 40 and 250 Myrs, then the radiative zone disappears and stars become fully convective throughout their life. For M dwarfs with masses less than 0.3$M_{\odot}$, no radiative core has ever been formed, so there is no additional angular momentum reservoir in their entire life. Consequently, we speculate that the activity dip could also relate to the disappearance of the radiative core for potential young stars with masses between 0.3 and 0.32$M_{\odot}$ as they evolve. However, the connection to the $^3He$ instability encounters two challenges: 1) stars with such a young age in this mass range would mostly be elevated higher than the main sequence, so they won't be included in our analysis, and 2) this age range is much younger than the spin-down age determined empirically by \cite{Pass2022}.

\section{Summary}
\label{sec:conclusion}

We carefully select and observe 480 M dwarfs on either side of the gap to understand how the interior instability may affect their H$\alpha$ activities. Our high-resolution spectroscopic survey results, the largest high-resolution survey focusing on this tight transition zone, show the H$\alpha$ emission is almost not detectable for stars in the bottom half of our ROI, with a few exceptions close to the red edge of the ROI based on results in J18 and Z21. This implies that stars in the gap don't show excessive H$\alpha$ emissions compared to stars above and below the gap, and observationally the H$\alpha$ activity has a sharp transition at the GE. Because the gap marks the interior transition, those active stars are mostly partially convective, and stars in and below the gap are mostly inactive. This relatively sharp activity transition is different from results from a large survey of Ly$\alpha$ and X-ray of M dwarfs by \cite{Linsky2020}, where they found no abrupt changes at or near the boundary in Ly$\alpha$ and X-ray fluxes. 

Our targets have relatively consistent ages and metallicities based on studies of CaH1 strength and galactic kinematics, but their true ages or metallicities may still differ. Hence, we still have heterogeneous stars within the ROI. As we know that the $^{3}$He instability can make their interior structure change depending on their ages, so we should have a wide variety of interior structures for stars in the gap. Yet, their interior differences are not reflected in the H$\alpha$ activities.

Furthermore, the majority of active stars are in the top half of our
ROI and are also above the best-fitted main sequence or in region 2 in
Figure~\ref{fig:Newton}. This slightly shifted distribution of active
stars can also be seen in N17, J18, and Z21. Based on a large number
of low and median resolution spectra from LAMOST in Z21, we find that
active stars are shifted up to $\sim$0.07 mag redder in
$G_{BP}-G_{RP}$ for stars at $M_{G}=$10, but the distribution of
active stars fainter than $M_{G}\sim$11 generally matches the
curvature of the main sequence. Key to this discussion is that
this distribution of active stars on HRD is similar to the
distribution of fast-rotation stars shown in
Figure~\ref{fig:Reinhold}. Also relevant to the link between M dwarf
rotation and activity is the study by \cite{West2015}, who reported
that the H$\alpha$ activity in 238 nearby M dwarfs is closely
associated with rotation. Here, we provide further evidence of a
strong correlation between stellar activity and rotation on the 2D
HRD.

The rationale behind the design of the ROI is to encompass stars on
both sides of the gap rather than the main sequence. As a result, the
bottom half of the ROI primarily consists of stars situated on the
blue side of the main sequence. Essentially, the absence of active
stars in the bottom half of the ROI could suggest that fast rotators
may have shifted redward outside of the ROI and above the main
sequence and that stars left in the bottom half of the ROI are mostly
slow rotators. If the internal instability were capable of triggering
additional magnetic activities on the stellar surface, we would have
detected excess H$\alpha$ activities within the magnitude range of
10$<M_{G}<$10.3, in addition to the fast rotators. However,
Figure~\ref{fig:newfeature} does not show any anomaly in H$\alpha$
activity within this magnitude range, indicating that the fast
rotators primarily trigger the detected H$\alpha$
emission. Furthermore, both this study and previous investigations on
H$\alpha$ emission have detected a minimal number of stars in emission
in the bottom half of the ROI, especially among stars within the
gap. This suggests that either the internal instability in these M
dwarfs has minimal impact on atmospheric activities, and/or that any
potential influence from the internal instability does not manifest as
H$\alpha$ emission.

Fast rotators often have spots, and the starspot model by \cite{Somers2020} suggested that starspots could make low-mass stars in the Pleiades redder depending on the spot coverage. However, the presence of starspots doesn't always mean the presence of an H$\alpha$ emission. For example, the solar spectrum has an H$\alpha$ absorption \citep{Wallace2011}, but still has spots. The formation mechanism of the H$\alpha$ line in M dwarf atmosphere is intertwined between rotation, chromospheric density, magnetic activities, and photosphere temperature \citep{Basri2021}. We don't know whether all these active stars flagged by this work and others have starspots present on their surfaces. Even if all these active stars have starspots, that will imply most stars have a similar spot coverage to make these active stars consistently shift redder. On the other hand, \cite{Jackson2014} showed an opposite shift in colors for stars with starspots. So how the starspots affect their colors is not clear, but our result seems to match the model proposed by \cite{Somers2020} if the cause of the color shift is starspots.

For stars below the gap with $M_{G}>11$, active fully convective stars generally follow the curvature of the main sequence, and the shift disappears. If we adopt the model by \cite{Somers2020} that starspots can make partially convective stars redder, the presence of starspots on these fully convective stars would appear to have minimum or no impact on their $G_{BP}-G_{RP}$ colors. \cite{Berdyugina2005} also showed the spot temperature contrast with respect to the photospheric temperature is less than 500 K for dwarfs cooler than 3500 K. Hence, the starspot's contrast on the photosphere would have minimal effect on the effective temperature, so $G_{BP}-G_{RP}$ color differences between active and inactive stars appear to be indistinguishable for stars fainter than $M_{G}\sim$11.

Finally, while we study the H$\alpha$ activities using the LAMOST data, we unveil some stars at 10.3$<M_{G}<$10.8 lack H$\alpha$ emissions. The percentage of active stars drops a few percent within this magnitude range. This $M_{G}$ region is also overlapped with the bottom half of our ROI, where only one active star is identified from this work. Z21 discussed that the activity dip might be caused by changing the magnetic dynamo mechanism at the boundary, but our work shows the activity dip is below the gap. Based on the large number of rotation periods presented in RH20, we found this activity dip coincides with a region where fewer stars rotate faster than $P_{crit}$ periods. This implies some of the most massive fully convective stars spin down rapidly, so fewer stars become active. However, the link between H$\alpha$ activity dip and rotation dip is based on the current large survey of nearby M dwarfs from the K2 mission. We lack all rotation periods for unbiased and completed late M dwarfs. Only after both their rotation and H$\alpha$ lines are measured can we further understand and explain this newly identified H$\alpha$ activity dip on the lower main sequence.

\begin{acknowledgments}
We would like to thank Liyun Zhang, Zhong-Rui Bai, and Luqian Wang for their help with the LAMOST data. We also thank Emily Pass and Gregory Feiden for their suggestions. This work was supported by the NASA Astrophysics Data Analysis Program (ADAP) under grant 20-ADAP20-0288. We have used data from the CHIRON spectrograph on the SMARTS 1.5m telescope, which is operated as part of the SMARTS Consortium by RECONS (www.recons.org) members, and with the assistance of staff, especially Roberto Aviles, and Rodrigo Hinojosa, at Cerro Tololo Inter-American Observatory.

This research has made use of the SIMBAD database, operated at CDS, Strasbourg, France. This work has made use of data from the European Space Agency (ESA) mission {\it Gaia} (\url{https://www.cosmos.esa.int/gaia}), processed by the {\it Gaia} Data Processing and Analysis Consortium (DPAC, \url{https://www.cosmos.esa.int/web/gaia/dpac/consortium}).  Funding for the DPAC has been provided by national institutions, in particular the institutions participating in the {\it Gaia} Multilateral Agreement. Guoshoujing Telescope (the Large Sky Area Multi-Object Fiber Spectroscopic Telescope LAMOST) is a National Major Scientific Project built by the Chinese Academy of Sciences. Funding for the project has been provided by the National Development and Reform Commission. LAMOST is operated and managed by the National Astronomical Observatories, Chinese Academy of Sciences. This research has made use of NASA's Astrophysics Data System.

\end{acknowledgments}

\facilities{CTIO:1.5m (CHIRON)}

\software{Matplotlib
  \citep{Hunter2007},
  NumPy \citep{vanderWalt2011}, astropy \citep{Astropy2013, Astropy2018},  SciPy \citep{Virtanen2020} and TOPCAT \citep{TOPCAT}}

\appendix

\section{Active and inactive Stars in Z21}
\label{sec:appendex1}

\begin{figure}
    \centering
    \plotone{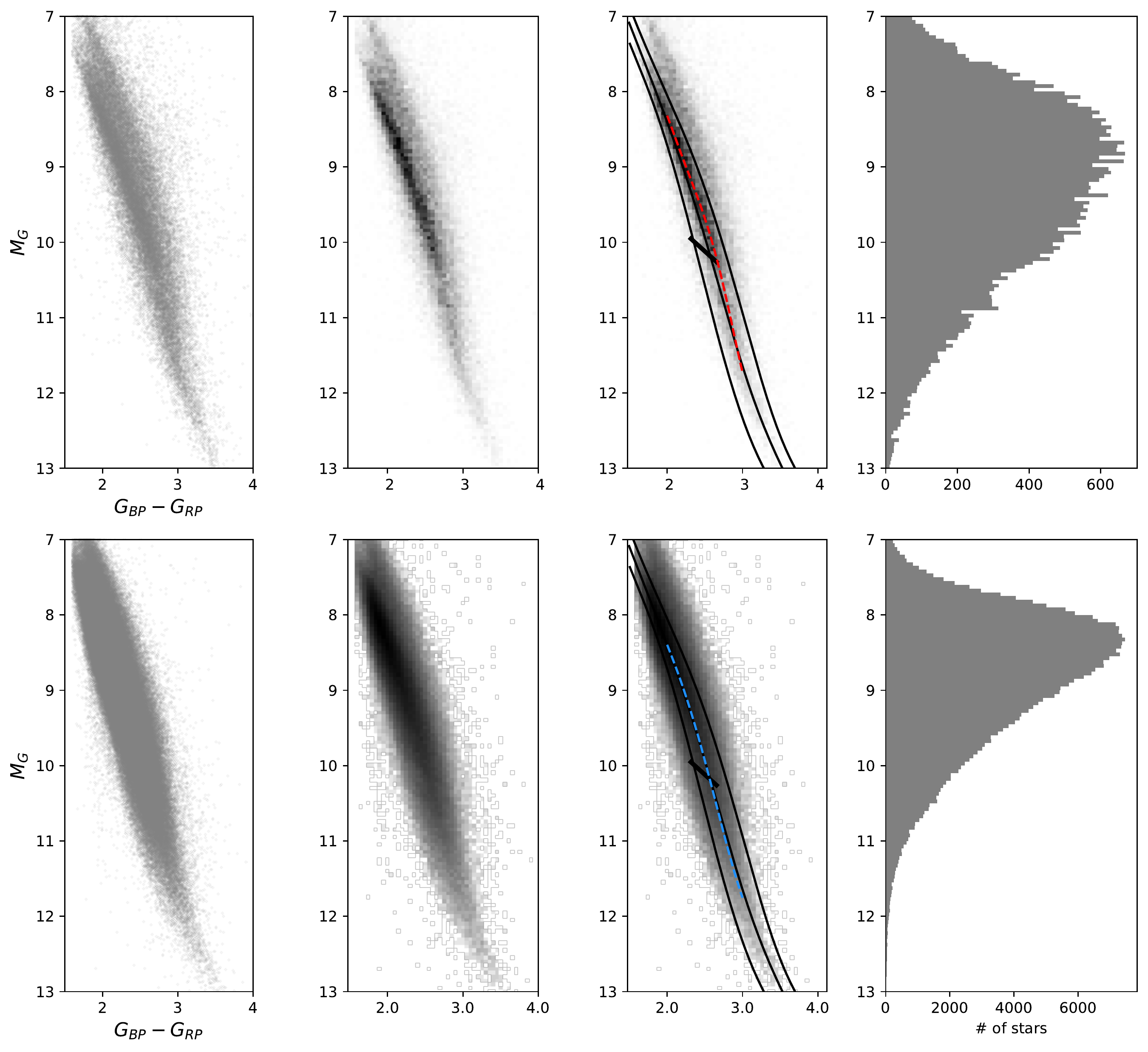}
    \caption{This figure is the same as Figure~\ref{fig:Zhangtight}, but all active stars flagged by Z21 are shown. The histogram dip at $M_{G}\sim$10.5 is still seen. In contrast, the histogram of inactive stars continues to be smooth as the distribution seen in Figure~\ref{fig:Zhangtight}. This shows identifying this dip is not related to how the active/inactive stars are selected.}
    \label{fig:LAMOSTall}
\end{figure}

Without using our new H$\alpha$ selecting criteria discussed in section~\ref{sec:LAMOST}, the entire active and inactive stars flagged by Z21 in the HRD are shown in Figure~\ref{fig:LAMOSTall}. The dip for active stars at $M_{G}$ 10.5 in the histogram is also seen. This demonstrates no matter what criteria are used, the percentage of active stars at $M_{G}$ 10.5 is low. The best-fitted distributions for active and inactive stars are almost identical to the best-fitted distributions in Figure~\ref{fig:LAMOSTall}. Averagely, active stars above the gap are 0.05 mag redder than the best-fitted main sequence. 

\section{A few additional large surveys}
Two additional large H$\alpha$ surveys for M dwarfs and one work with a large number of rotation periods discussed next are excluded from our analysis. 

\subsection{Kiman 2019}
\label{sec:Kidman2019}

\begin{figure}
    \centering
    \includegraphics[scale=0.7]{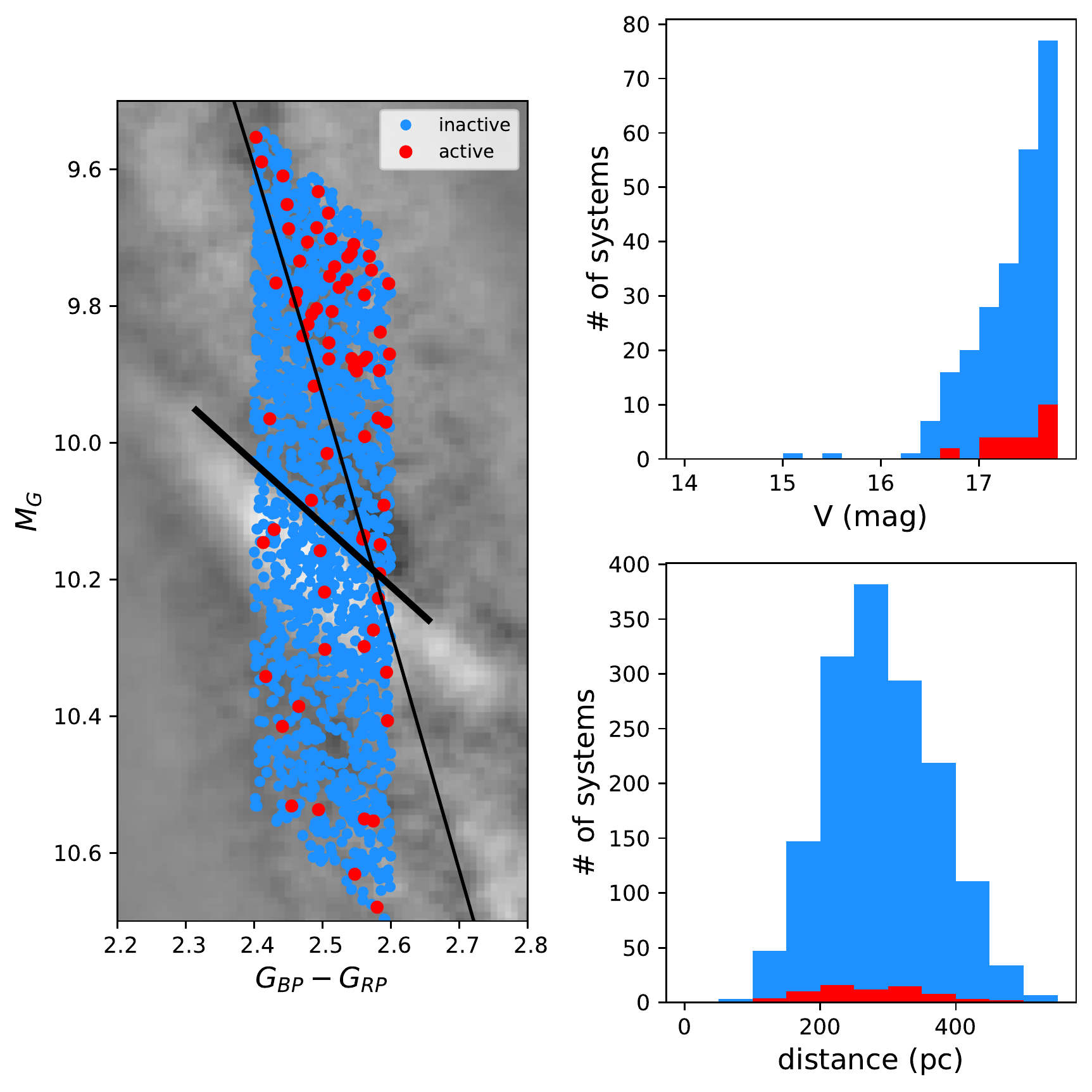}
    \caption{The left figure shows active (red dots) and inactive (blue dots) stars reported in \cite{Kiman2019} in our ROI. The top right and the bottom right figures are histograms of estimated V magnitudes and distances respectively for stars shown in the left figure. The red and blue histograms are for active and inactive stars, respectively.}
    \label{fig:SDSS}
\end{figure}

\citet{Kiman2019}, hereafter K19, reported H$\alpha$ activities of more than 74,000 M and L dwarfs by matching targets in SDSS and Gaia DR2. However, this large sample has a mean parallax of 3.77 mas or a mean distance of 265 pc. There are only 3 stars with distances less than 100 pc in our ROI after we apply the following selecting criteria: 1) RUWE $<$1.4, 2) photometric\_sample\_subred $=$1, 3) astrometric\_sample $=$1, 4) $\pi/\pi_{error}>$10, 5) no Gaia EDR3 source of $\Delta G_{RP}<$4.0 within the 4$^{\prime\prime}$, where the SDSS fiber has a 3$^{\prime\prime}$ diameter.  In total, there are 1,671 in this region, and 71 stars are flagged active by K19. Compared to the samples in N17, J18, Z21, and this work, the K19 sample could potentially include more unresolved close binaries because the mean distance is far, and almost all stars are fainter than $V=$15 mag. Nonetheless, all active and inactive stars within our ROI are shown in Figure~\ref{fig:SDSS}, and K19 has significantly more active stars in the bottom half of this ROI. We think because the distance cutoff is different from others, the results in K19 are different than others. Besides, their sample includes M dwarfs selected from \cite{West2011}, where they had a non-uniform sample selection, so obvious sampling gaps can be seen in their color-magnitude Diagram and HRD \citep{West2011, Kiman2019}. Hence, it is difficult to apply their results to understand stars in the ROI.

\subsection{Anthony 2022}
\cite{Anthony2022} reported H$\alpha$'s EW of 122 M dwarfs in the Southern Continuous Viewing Zone (CVZ) of the TESS mission using low-resolution spectra, and the mean parallax is 25.9 mas. Only 14 targets are within our ROI. A total of 12 stars with H$\alpha<-$0.5\AA~are identified as active, but no active M dwarfs are within our ROI. Seven of the 12 active stars are fully convective stars. 

\subsection{Popinchalk 2021}
\label{app:Popinchalk}

\begin{figure}
    \centering
    \includegraphics[scale=0.7]{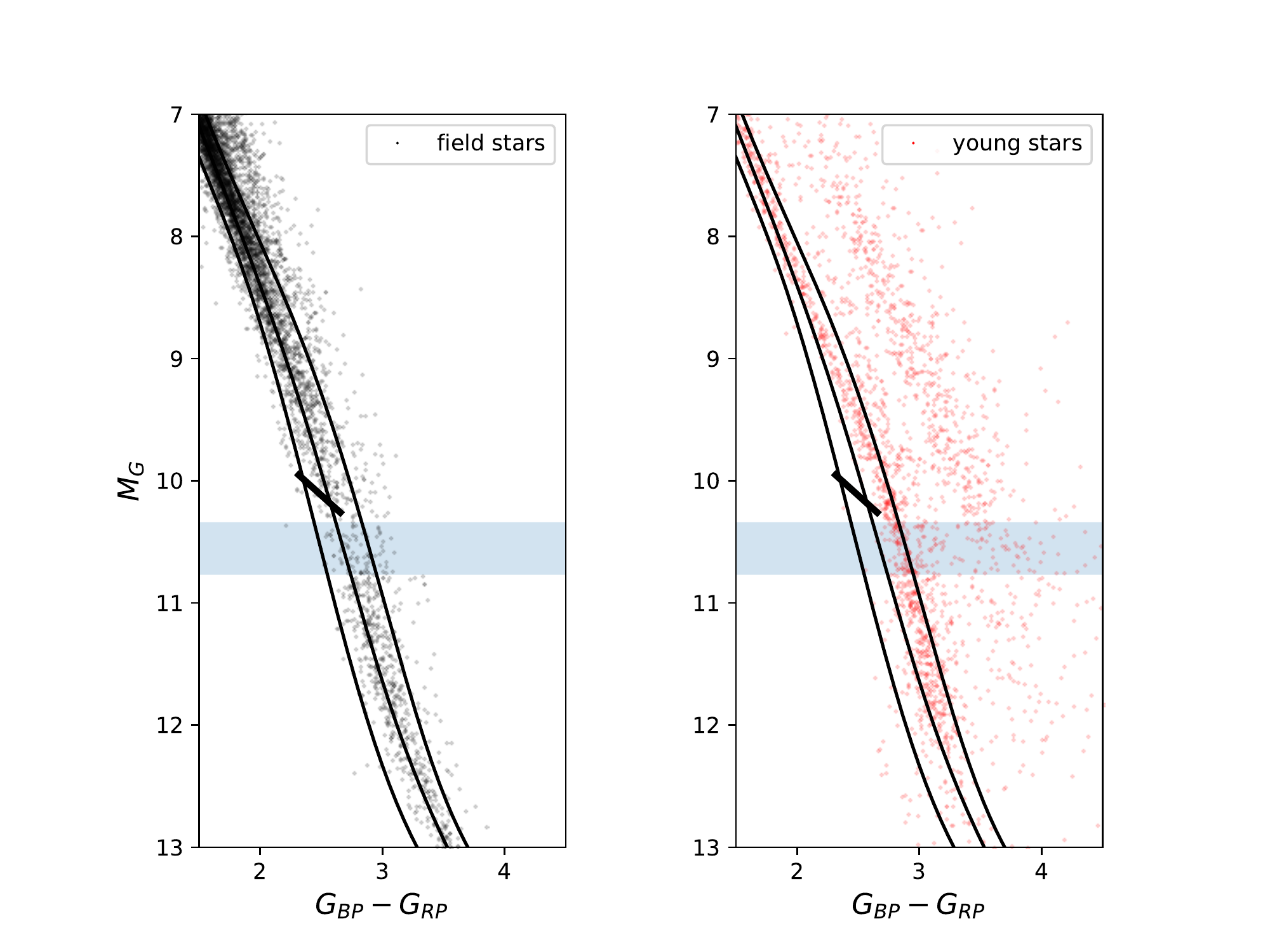}
    \caption{Field stars (black dots in the left plot) and young stars (red dots in the right plot) in clusters with rotational periods from \cite{Popinchalk2021}. The blue boxes indicate the activity dip in Figure~\ref{fig:newfeature}.}
    \label{fig:Popinchalk}
\end{figure}

\cite{Popinchalk2021} presented 8,296 rotational periods for field stars and stars in young clusters with various ages in their Table 3. The distributions of these two populations on the HRD are shown in Figure~\ref{fig:Popinchalk}. We can see that the number of field stars within the highlighted box is noticeably lower than the stars above and below this box. Thus, they don't have enough stars with this particular region to conduct statistics. As for young stars, the "main sequences" of different ages of clusters don't follow the main sequence of nearby stars. The lower band of the "main sequence" is mainly composed of stars from Pleiades, Praesepe, and Hyades, and the upper band is mainly from stars in the Upper-Sco moving group. Fields stars below the gap may have been selected from various biases, so the star density is low. Therefore, stars are undersampled in 10.3$<M_G<$10.8 in comparison with \cite{Reinhold2020} and \cite{Lu2022}.

\section{The Best fitted Main sequence}
\label{app:lines}

Here we present 1) the best-fitted top edge of the gap or GE, 2) the best-fitted main sequence, and 3) the upper and lower envelopes of the main sequences using data in Gaia EDR3. To generate these best-fitted lines, the algorithms and target selection used to generate the enhanced HRD discussed in \citet{Jao2020} are applied for stars in Gaia EDR3. In order to approximately use a fitted line to represent the two-dimensional gap, we fit a straight line to represent the top edge of the gap shown in Figure~\ref{fig:results}. The upper and lower envelopes cover 90\% of the populations at a given $G_{BP}-G_{RP}$ color using a skewed Gaussian curve to represent the population. The peak of the distribution indicates the best-fitted main sequence at this given color. All the best-fitted polynomial coefficients for these lines are given in Table~\ref{tbl:coeffs}. 

\begin{deluxetable}{crrrr}
\tablecaption{Coefficients of the polynomials used to generate
  the best-fitted main sequence, the gap, and envelopes \label{tbl:coeffs}}
\tablehead{
\colhead{coeff.} &
\colhead{MS} &
\colhead{top of the gap} &
\colhead{upper envelope} &
\colhead{lower envelope}
}
\startdata
 a0 &   19.10849382 &   7.86323529 &  -13.58268031 &    13.12750985 \\
 a1 &  -95.44879765 &   0.90252101 &   57.62576454 &   -46.72259226 \\
 a2 &  214.54030704 &	           &  -79.81685633 &    90.64589118 \\
 a3 & -219.30493198 &	           &   64.86365682 &   -82.39866903 \\
 a4 &   87.35269064 &	           &  -31.37744697 &    40.94150742 \\
 a5 &   35.52596873 &	           &    8.88026358 &   -11.26957046 \\
 a6 &  -61.88803250 &	           &   -1.34570736 &     1.61207258 \\
 a7 &   34.78644351 &	           &    0.08384474 &    -0.09356802 \\
 a8 &  -10.88510027 &	           &		       & 		        \\
 a9 &    2.00794707 &	           &		       &		        \\
a10 &   -0.20466398 &	           &		       &		        \\
a11 &    0.00891475 &	           &		       &		        \\
\enddata
\tablecomments{The relations are valid at
  0.8$\leq$$G_{BP}-G_{RP}$$\leq$4.0 for the MS and two envelopes, and 
  2.3$\leq$$G_{BP}-G_{RP}$$\leq$2.9 for the gap. All relations have the format of $M_{G}=\sum_{n=0}^{order} a_n\times(G_{BP}-G_{RP})^{n}$.}
\end{deluxetable}


\begin{thebibliography}{}

\bibitem[Anthony et al.(2022)]{Anthony2022} Anthony, F., N{\'u}{\~n}ez, A., Ag{\"u}eros, M.~A., et al.\ 2022, \aj, 163, 257. doi:10.3847/1538-3881/ac6110

\bibitem[Astropy Collaboration et al.(2013)]{Astropy2013} Astropy Collaboration, Robitaille, T.~P., Tollerud, E.~J., et al.\ 2013, \aap, 558, A33. doi:10.1051/0004-6361/201322068

\bibitem[Astropy Collaboration et al.(2018)]{Astropy2018} Astropy Collaboration, Price-Whelan, A.~M., Sip{\H{o}}cz, B.~M., et al.\ 2018, \aj, 156, 123. doi:10.3847/1538-3881/aabc4f

\bibitem[Bailey et al.(2012)]{Bailey2012} Bailey, J.~I., White, R.~J., Blake, C.~H., et al.\ 2012, \apj, 749, 16. doi:10.1088/0004-637X/749/1/16

\bibitem[Baraffe \& Chabrier(2018)]{Baraffe2018} Baraffe, I. \& Chabrier, G.\ 2018, \aap, 619, A177. doi:10.1051/0004-6361/201834062

\bibitem[Basri(2021)]{Basri2021} Basri, G.\ 2021, An Introduction to Stellar Magnetic Activity, by Basri, Gibor. ISBN: 978-0-7503-2130-3. IOP ebooks. Bristol, UK: IOP Publishing, 2021. doi:10.1088/2514-3433/ac2956

\bibitem[Berdyugina(2005)]{Berdyugina2005} Berdyugina, S.~V.\ 2005, Living Reviews in Solar Physics, 2, 8. doi:10.12942/lrsp-2005-8

\bibitem[Bessell(1982)]{Bessell1982} Bessell, M.~S.\ 1982, \pasa, 4, 417. doi:10.1017/S1323358000021329

\bibitem[Bochanski et al.(2007)]{Bochanski2007} Bochanski, J.~J., West, A.~A., Hawley, S.~L., et al.\ 2007, \aj, 133, 531. doi:10.1086/510240

\bibitem[Burgasser et al.(2003)]{Burgasser2003} Burgasser, A.~J., Kirkpatrick, J.~D., Reid, I.~N., et al.\ 2003, \apj, 586, 512. doi:10.1086/346263

\bibitem[Cayrel(1988)]{Cayrel1988} Cayrel, R.\ 1988, The Impact of Very High S/N Spectroscopy on Stellar Physics, 132, 345

\bibitem[Charbonneau(2014)]{Charbonneau2014} Charbonneau, P.\ 2014, \araa, 52, 251. doi:10.1146/annurev-astro-081913-040012

\bibitem[Chabrier \& K{\"u}ker(2006)]{Chabrier2006} Chabrier, G. \& K{\"u}ker, M.\ 2006, \aap, 446, 1027. doi:10.1051/0004-6361:20042475

\bibitem[Clements et al.(2017)]{Clements2017} Clements, T.~D., Henry, T.~J., Hosey, A.~D., et al.\ 2017, \aj, 154, 124. doi:10.3847/1538-3881/aa8464

\bibitem[Cram \& Mullan(1979)]{Cram1979} Cram, L.~E. \& Mullan, D.~J.\ 1979, \apj, 234, 579. doi:10.1086/157532

\bibitem[Curtis et al.(2020)]{Curtis2020} Curtis, J.~L., Ag{\"u}eros, M.~A., Matt, S.~P., et al.\ 2020, \apj, 904, 140. doi:10.3847/1538-4357/abbf58

\bibitem[Donati et al.(2008)]{Donati2008} Donati, J.-F., Morin, J., Petit, P., et al.\ 2008, \mnras, 390, 545. doi:10.1111/j.1365-2966.2008.13799.x

\bibitem[Douglas et al.(2014)]{Douglas2014} Douglas, S.~T., Ag{\"u}eros, M.~A., Covey, K.~R., et al.\ 2014, \apj, 795, 161. doi:10.1088/0004-637X/795/2/161

\bibitem[Dungee et al.(2022)]{Dungee2022} Dungee, R., van Saders, J., Gaidos, E., et al.\ 2022, \apj, 938, 118. doi:10.3847/1538-4357/ac90be

\bibitem[Feiden et al.(2021)]{Feiden2021} Feiden, G.~A., Skidmore, K., \& Jao, W.-C.\ 2021, \apj, 907, 53. doi:10.3847/1538-4357/abcc03

\bibitem[Fouqu{\'e} et al.(2018)]{Fouque2018} Fouqu{\'e}, P., Moutou, C., Malo, L., et al.\ 2018, \mnras, 475, 1960. doi:10.1093/mnras/stx3246

\bibitem[Gaia Collaboration et al.(2021)]{EDR3} Gaia Collaboration, Brown, A.~G.~A., Vallenari, A., et al.\ 2021, \aap, 649, A1. doi:10.1051/0004-6361/202039657

\bibitem[Gaia Collaboration et al.(2022)]{nonsingle2022} Gaia Collaboration, Arenou, F., Babusiaux, C., et al.\ 2022, arXiv:2206.05595

\bibitem[Gilman(2005)]{Gilman2005} Gilman, P.~A.\ 2005, Astronomische Nachrichten, 326, 208. doi:10.1002/asna.200410378

\bibitem[Gizis(1997)]{Gizis1997} Gizis, J.~E.\ 1997, \aj, 113, 806. doi:10.1086/118302

\bibitem[Gizis et al.(2002)]{Gizis2002} Gizis, J.~E., Reid, I.~N., \& Hawley, S.~L.\ 2002, \aj, 123, 3356. doi:10.1086/340465

\bibitem[Irwin et al.(2018)]{Irwin2018} Irwin, J.~M., Charbonneau, D., Esquerdo, G.~A., et al.\ 2018, \aj, 156, 140. doi:10.3847/1538-3881/aad9a3

\bibitem[Hawley et al.(1996)]{Hawley1996} Hawley, S.~L., Gizis, J.~E., \& Reid, I.~N.\ 1996, \aj, 112, 2799. doi:10.1086/118222

\bibitem[Henry et al.(1994)]{Henry1994} Henry, T.~J., Kirkpatrick, J.~D., \& Simons, D.~A.\ 1994, \aj, 108, 1437. doi:10.1086/117167

\bibitem[Houdebine \& Mullan(2015)]{Houdebine2015} Houdebine, E.~R. \& Mullan, D.~J.\ 2015, \apj, 801, 106. doi:10.1088/0004-637X/801/2/106

\bibitem[Houdebine et al.(2017)]{Houdebine2017} Houdebine, E.~R., Mullan, D.~J., Bercu, B., et al.\ 2017, \apj, 837, 96. doi:10.3847/1538-4357/aa5cad

\bibitem[Hubbard-James et al.(2022)]{James2022} Hubbard-James, H.S. et al.\ 2022, submitted to AJ

\bibitem[Hunter (2007)]{Hunter2007} Hunter, J.D., \ 2007, Computing in Science \& Engineering, 9, 90

\bibitem[Jackson \& Jeffries(2014)]{Jackson2014} Jackson, R.~J. \& Jeffries, R.~D.\ 2014, \mnras, 445, 4306. doi:10.1093/mnras/stu2076

\bibitem[Jao et al.(2011)]{Jao2011} Jao, W.-C., Henry, T.~J., Subasavage, J.~P., et al.\ 2011, \aj, 141, 117. doi:10.1088/0004-6256/141/4/117

\bibitem[Jao et al.(2018)]{Jao2018} Jao, W.-C., Henry, T.~J., Gies,
  D.~R., \& Hambly, N.~C.\ 2018, \apjl, 861, L11

\bibitem[Jao \& Feiden(2020)]{Jao2020} Jao, W.-C. \& Feiden, G.~A.\ 2020, \aj, 160, 102. doi:10.3847/1538-3881/aba192

\bibitem[Jao \& Feiden(2021)]{Jao2021} Jao, W.-C. \& Feiden, G.~A.\ 2021, Research Notes of the American Astronomical Society, 5, 124. doi:10.3847/2515-5172/ac053a

\bibitem[Jao et al.(2022)]{Jao2022} Jao, W.-C., Couperus, A.~A., Vrijmoet, E.~H., et al.\ 2022, \apj, 940, 145. doi:10.3847/1538-4357/ac9cd8

\bibitem[Jeffers et al.(2018)]{Jeffers2018} Jeffers, S.~V., Sch{\"o}fer, P., Lamert, A., et al.\ 2018, \aap, 614, A76. doi:10.1051/0004-6361/201629599

\bibitem[Linsky et al.(2020)]{Linsky2020} Linsky, J.~L., Wood, B.~E., Youngblood, A., et al.\ 2020, \apj, 902, 3. doi:10.3847/1538-4357/abb36f

\bibitem[Kiman et al.(2019)]{Kiman2019} Kiman, R., Schmidt, S.~J., Angus, R., et al.\ 2019, \aj, 157, 231. doi:10.3847/1538-3881/ab1753

\bibitem[Kiman et al.(2021)]{Kiman2021} Kiman, R., Faherty, J.~K., Cruz, K.~L., et al.\ 2021, \aj, 161, 277. doi:10.3847/1538-3881/abf561

\bibitem[Kirkpatrick et al.(1991)]{Kirkpatrick1991} Kirkpatrick, J.~D., Henry, T.~J., \& McCarthy, D.~W.\ 1991, \apjs, 77, 417. doi:10.1086/191611

\bibitem[Kraft(1967)]{Kraft1967} Kraft, R.~P.\ 1967, \apj, 150, 551. doi:10.1086/149359

\bibitem[Lindegren et al.(2018)]{Lindegren2018} Lindegren, L., Hern{\'a}ndez, J., Bombrun, A., et al.\ 2018, \aap, 616, A2. doi:10.1051/0004-6361/201832727

\bibitem[L{\'o}pez-Valdivia et al.(2021)]{Lopez2021} L{\'o}pez-Valdivia, R., Sokal, K.~R., Mace, G.~N., et al.\ 2021, \apj, 921, 53. doi:10.3847/1538-4357/ac1a7b

\bibitem[Lu et al.(2022)]{Lu2022} Lu, Y.~L., Curtis, J.~L., Angus, R., et al.\ 2022, \aj, 164, 251. doi:10.3847/1538-3881/ac9bee

\bibitem[MacDonald \& Gizis(2018)]{MacDonald2018} MacDonald, J. \& Gizis, J.\ 2018, \mnras, 480, 1711. doi:10.1093/mnras/sty1888

\bibitem[Malo et al.(2014)]{Malo2014} Malo, L., Artigau, {\'E}., Doyon, R., et al.\ 2014, \apj, 788, 81. doi:10.1088/0004-637X/788/1/81

\bibitem[Medina et al.(2022)]{Medina2022} Medina, A.~A., Winters, J.~G., Irwin, J.~M., et al.\ 2022, \apj, 935, 104. doi:10.3847/1538-4357/ac77f9

\bibitem[Pecaut \& Mamajek(2013)]{Pecaut2013} Pecaut, M.~J. \& Mamajek, E.~E.\ 2013, \apjs, 208, 9. doi:10.1088/0067-0049/208/1/9

\bibitem[Mansfield \& Kroupa(2021)]{Mansfield2021} Mansfield, S. \& Kroupa, P.\ 2021, \aap, 650, A184. doi:10.1051/0004-6361/202140536

\bibitem[Mohanty \& Basri(2003)]{Mohanty2003} Mohanty, S. \& Basri, G.\ 2003, \apj, 583, 451. doi:10.1086/345097

\bibitem[Namekata et al.(2021)]{Namekata2022} Namekata, K., Maehara, H., Honda, S., et al.\ 2021, Nature Astronomy, 6, 241. doi:10.1038/s41550-021-01532-8

\bibitem[Newton et al.(2017)]{Newton2017} Newton, E.~R., Irwin, J., Charbonneau, D., et al.\ 2017, \apj, 834, 85. doi:10.3847/1538-4357/834/1/85

\bibitem[Nisak et al.(2022)]{Nisak2022} Nisak, A.~H., White, R.~J., Yep, A., et al.\ 2022, \aj, 163, 278. doi:10.3847/1538-3881/ac63c3

\bibitem[Paredes et al.(2021)]{Paredes2021} Paredes, L.~A., Henry, T.~J., Quinn, S.~N., et al.\ 2021, \aj, 162, 176. doi:10.3847/1538-3881/ac082a

\bibitem[Passegger et al.(2018)]{Passegger2018} Passegger, V.~M., Reiners, A., Jeffers, S.~V., et al.\ 2018, \aap, 615, A6. doi:10.1051/0004-6361/201732312

\bibitem[Pass et al.(2022)]{Pass2022} Pass, E.~K., Charbonneau, D., Irwin, J.~M., et al.\ 2022, \apj, 936, 109. doi:10.3847/1538-4357/ac7da8

\bibitem[Popinchalk et al.(2021)]{Popinchalk2021} Popinchalk, M., Faherty, J.~K., Kiman, R., et al.\ 2021, \apj, 916, 77. doi:10.3847/1538-4357/ac0444

\bibitem[van Saders \& Pinsonneault(2012)]{Saders2012} van Saders, J.~L. \& Pinsonneault, M.~H.\ 2012, \apj, 751, 98. doi:10.1088/0004-637X/751/2/98

\bibitem[Reid et al.(1995)]{Reid1995} Reid, I.~N., Hawley, S.~L., \& Gizis, J.~E.\ 1995, \aj, 110, 1838. doi:10.1086/117655

\bibitem[Reiners \& Basri(2009)]{Reiners2009} Reiners, A. \& Basri, G.\ 2009, \aap, 496, 787. doi:10.1051/0004-6361:200811450

\bibitem[Reiners et al.(2014)]{Reiners2014} Reiners, A., Sch{\"u}ssler, M., \& Passegger, V.~M.\ 2014, \apj, 794, 144. doi:10.1088/0004-637X/794/2/144

\bibitem[Reiners et al.(2018)]{Reiners2018} Reiners, A., Zechmeister, M., Caballero, J.~A., et al.\ 2018, \aap, 612, A49. doi:10.1051/0004-6361/201732054

\bibitem[Reiners et al.(2022)]{Reiners2022} Reiners, A., Shulyak, D., K{\"a}pyl{\"a}, P.~J., et al.\ 2022, \aap, 662, A41. doi:10.1051/0004-6361/202243251

\bibitem[Reinhold \& Hekker(2020)]{Reinhold2020} Reinhold, T. \& Hekker, S.\ 2020, \aap, 635, A43. doi:10.1051/0004-6361/201936887

\bibitem[Scholz et al.(2007)]{Scholz2007} Scholz, A., Coffey, J., Brandeker, A., et al.\ 2007, \apj, 662, 1254. doi:10.1086/518361

\bibitem[Sch{\"o}nrich et al.(2010)]{Schonrich2010} Sch{\"o}nrich, R., Binney, J., \& Dehnen, W.\ 2010, \mnras, 403, 1829. doi:10.1111/j.1365-2966.2010.16253.x

\bibitem[Soderblom(1982)]{Soderblom1982} Soderblom, D.~R.\ 1982, \apj, 263, 239. doi:10.1086/160498

\bibitem[Somers et al.(2020)]{Somers2020} Somers, G., Cao, L., \& Pinsonneault, M.~H.\ 2020, \apj, 891, 29. doi:10.3847/1538-4357/ab722e

\bibitem[Spada \& Lanzafame(2020)]{Spada2020} Spada, F. \& Lanzafame, A.~C.\ 2020, \aap, 636, A76. doi:10.1051/0004-6361/201936384

\bibitem[Stepien(1994)]{Stepien1994} Stepien, K.\ 1994, \aap, 292, 191

\bibitem[Taylor(2005)]{TOPCAT} Taylor, M.~B.\ 2005, Astronomical Data Analysis Software and Systems XIV, 347, 29

\bibitem[Tokovinin et al.(2013)]{Tokovinin2013} Tokovinin, A., Fischer, D.~A., Bonati, M., et al.\ 2013, \pasp, 125, 1336. doi:10.1086/674012


\bibitem[Van der Walt, Colbert \& Varoquaux (2011)]{vanderWalt2011} van der Walt, S., Colbert, S.C., Varoquaux, G., \ 2011, Computing in Science
  \& Engineering, 13, 22
  
\bibitem[Virtanen et. al. (2020)]{Virtanen2020} Virtanen, P., Gommers, R., Oliphant, T.E., et al., \ 2020, Nature Methods, 17, 261

\bibitem[Walkowicz et al.(2004)]{Walkowicz2004} Walkowicz, L.~M., Hawley, S.~L., \& West, A.~A.\ 2004, \pasp, 116, 1105. doi:10.1086/426792

\bibitem[Walkowicz \& Hawley(2009)]{Walkowicz2009} Walkowicz, L.~M. \& Hawley, S.~L.\ 2009, \aj, 137, 3297. doi:10.1088/0004-6256/137/2/3297

\bibitem[Wallace et al.(2011)]{Wallace2011} Wallace, L., Hinkle, K.~H., Livingston, W.~C., et al.\ 2011, \apjs, 195, 6. doi:10.1088/0067-0049/195/1/6

\bibitem[West et al.(2008)]{West2008} West, A.~A., Hawley, S.~L., Bochanski, J.~J., et al.\ 2008, \aj, 135, 785. doi:10.1088/0004-6256/135/3/785

\bibitem[West et al.(2011)]{West2011} West, A.~A., Morgan, D.~P., Bochanski, J.~J., et al.\ 2011, \aj, 141, 97. doi:10.1088/0004-6256/141/3/97

\bibitem[White \& Hillenbrand(2004)]{White2004} White, R.~J. \& Hillenbrand, L.~A.\ 2004, \apj, 616, 998. doi:10.1086/425115

\bibitem[West et al.(2015)]{West2015} West, A.~A., Weisenburger, K.~L., Irwin, J., et al.\ 2015, \apj, 812, 3. doi:10.1088/0004-637X/812/1/3

\bibitem[Wright \& Eastman(2014)]{Wright2014} Wright, J.~T. \& Eastman, J.~D.\ 2014, \pasp, 126, 838. doi:10.1086/678541

\bibitem[Wright et al.(2018)]{Wright2018} Wright, N.~J., Newton, E.~R., Williams, P.~K.~G., et al.\ 2018, \mnras, 479, 2351. doi:10.1093/mnras/sty1670

\bibitem[Zhang et al.(2021)]{Zhang2021} Zhang, L.-Y., Meng, G., Long, L., et al.\ 2021, \apjs, 253, 19. doi:10.3847/1538-4365/abd7a8

\bibitem[Zhang et al.(2021)]{Zhang2021b} Zhang, S., Luo, A.-L., Comte, G., et al.\ 2021, \apj, 908, 131. doi:10.3847/1538-4357/abcfc5

\end{thebibliography}
\end{document}